\documentclass[a4paper,10pt]{article}
\usepackage{amsthm,amsmath}
\usepackage{amssymb}
\usepackage{fullpage}
\usepackage[utf8]{inputenc} 
\usepackage{graphicx}
\usepackage{version}
\usepackage{lipsum}
\usepackage{fixltx2e}
\usepackage{mathptmx}
\usepackage{times}
\usepackage{sidecap}

\newcommand{\hide}[1]{}

\newtheorem{thm}{Theorem}[section]
\newtheorem{claim}[thm]{Claim}
\newtheorem{lem}[thm]{Lemma}
\newtheorem{define}[thm]{Definition}

\def \Ed{E_{d}}
\def \Etheta{E_{\theta}}
\def \Ephi{E_{\varphi}}
\def \Evdw{E_{\textnormal{vdw}}}
\def \Ecoul{E_{\textnormal{coul}}}

\def \Gsol{E_{\textnormal{sol}}}
\def \Gpol{E_{\textnormal{pol}}}
\def \Gcav{E_{\textnormal{cav}}}
\def \Gvdw{E_{\textnormal{vdw(s-s)}}}

\newcommand{\bx}{\mathbf{x}}
\newcommand{\by}{\mathbf{y}}

\newcommand{\E}{\mathbf{E}}

\def \p{\partial}

\def \bvn{\vec{\bf n}}

\title{Quantifying and Visualizing Uncertainties in Molecular Models}

\author{Muhibur~Rasheed, Nathan~Clement, Abhishek~Bhowmick, and Chandrajit~Bajaj\\
Computer Science Department and Computational Visualization Center\\
The University of Texas at Austin, TX, 78712.\\
Corresponding Author: Chandrajit~Bajaj, bajaj@cs.utexas.edu.
}

\begin{document}

\maketitle

\begin{abstract}
Computational molecular modeling and visualization has seen significant progress in recent years with several molecular modeling and visualization software systems in use today. Nevertheless the molecular biology community lacks techniques and tools for the rigorous analysis, quantification and visualization of the associated errors in  molecular structure and its associated properties. This paper attempts at filling this vacuum with the introduction of  a systematic statistical framework where each source of structural uncertainty is modeled as a random variable (RV) with a known distribution, and properties of the molecules are defined as dependent RVs. The framework consists of a theoretical basis, and an empirical implementation where the uncertainty quantification (UQ) analysis is achieved by using Chernoff-like bounds. The framework enables additionally  the propagation of input structural data uncertainties, which in the molecular protein world are described as B-factors, saved with almost all X-ray models deposited in the Protein Data Bank (PDB). Our statistical framework is also able and has been applied to quantify and visualize the uncertainties in molecular properties, namely solvation interfaces and solvation free energy estimates. For each of these quantities of interest (QOI) of the molecular models we provide several novel and intuitive visualizations of the input, intermediate, and final propagated uncertainties. These methods should enable the end user achieve a more quantitative and visual evaluation of various molecular PDB models for structural and property correctness, or the lack thereof.
\end{abstract}

\section{Introduction}
Computational models of any biological substance are, by nature, prone to error. The source of this error can range anywhere from discrete representations of continuous data, to inadequate sampling of the parameter/search space, computational approximations, or even not sufficiently capturing all relevant aspects of the biological system. In some cases, these errors are slight or insignificant, but when the errors combine---as they frequently do for computations that involve geometry and complicated (linear or non-linear) numerical system---they can create a result that is unreliable. Modeling of protein structures and their interactions is a field of research that is especially susceptible to cascading errors, for various computed quantities of interest (QOIs). These computations include  multi-step methods for protein sequence alignment and homology modeling, implicit solvation interfaces (aka molecular surfaces) generation, configurationally dependent binding affinity calculations, 
molecular docking and structure refinement via molecular substructure replacement and fitting, etc. \cite{Richards77,Connolly83a,BLMP1997,BPS03,Eisenberg86,Nina97,Bashford00,feigbrooks04,Onufriev00,bajaj06f2dock,BajajZhao09,F2DockGBRerank}. For each and all of these computations related to molecular structure prediction, and energy estimations, the confidence in our results could necessarily be bolstered, if each estimation of a QOI in the computational pipeline also includes rigorous evaluation of their uncertainty. Such uncertainty bounds, when presented through intuitive visualization would be invaluable in rational design and analysis in molecular modeling tasks.

Consider the task of computationally predicting the correct folded structure of a protein and its binding interaction with a partner (e.g.\ an antibody or a small molecule (or ligand))- a common exercise is drug design. Typically a researcher would first try to identify if the structure is already available in the protein data bank (PDB) or the electron microscopy data bank (EMDB). While atomic resolution models (where locations of each atom is clearly reported as XYZ coordinates) exist for many proteins and protein-ligand complexes, in most real applications, one needs to resolve the structure for a new complex. Sometimes partial structures or structure for some components of an assembly is available, and the remaining parts need to be resolved either experimentally or computationally. Experimentally resolving such structures are time consuming, costly, and sometimes impossible due to the flexible nature of the molecules in question. In which case computational modeling is the only option.

\begin{figure*}[t!]
 \centering
\includegraphics[width=0.95\textwidth]{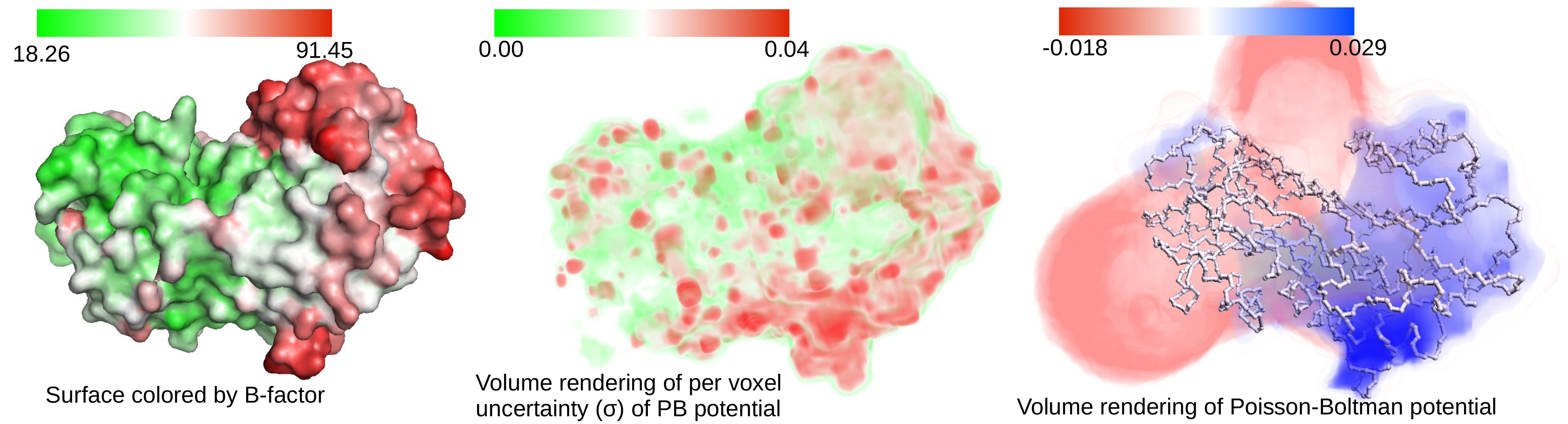}
\caption{An illustrative example of quantifying and visualizing uncertainties arising from noisy/uncertain input data. Atomic coordinates of molecules, determined using x-ray crystallography, comes with a measure of the uncertainty related to the coordinates. These measures, called B-factors, are positive numbers defined as $B=8\pi \langle u^2\rangle$, where $u$ defines the standard deviation of a Gaussian distribution which is centered at the reported atom location. Typically, B-factors above 50 are considered to represent high uncertainty. Current prediction and visualization techniques hardly incorporate the effect of such uncertainties into their results. For example, currently, one is able to see the distribution of high and low B-factors on the structural model of a molecule as we shown in (A), rendered using PyMol \cite{PyMOL}. Green-White-Red gradient is mapped to low-medium-high B-factors. The mathematical framework and visualization techniques introduced in this paper allows one to explore the effect of these uncertainties present in the input structure, on any computationally predicted properties or function. For instance the right panel of the figure shows the Poisson-Boltzman (PB) electrostatic potential inside and outside the same molecule (shown as backbone only). The potential is computed on each voxel of a volumetric map and rendered using blue (positive), white (neutral) and red (negative) colors. However, using our empirical uncertainty quantification model, one can (and should) further compute the uncertainty, expressed as the standard deviation $sigma$ computed across an ensemble of slightly perturbed samples of the molecule, of the potential computed at every voxel. Note that the standard deviation is quite high in some regions. This uncertainty is visualized in the middle panel. Notice that while the left and the middle figure has some correlation (high uncertainty regions tend to stay high uncertainty), it provides a more accurate and application dependent picture of effect the underlying uncertainty.}
\label{fig:pb-uq}
\end{figure*}

Such computational modeling tasks are typically expressed, in a mathematical sense, as an optimization problem over a high dimensional configuration space. The configuration space $\mathcal{C}$ is parameterized by the available degrees of the freedom for the molecules. We shall explore two separate parameterization, Cartesian coordinates and internal coordinates, in the methods section. For any given parameterization , the computational task involves sampling the space $\mathcal{C}$, evaluating a function $\mathcal{F}$ at each sample, and reporting the optimum of the function. The accuracy of the predicted optimum hinges upon a number of factors including the quality of the initial state or input structures to the system, the computational and numerical errors accumulated due to discretization and approximations adopted in the implementation of the scoring function, the dispersion and discrepancy of the sampling etc., resulting in a high degree of uncertainty on the predicted model/structure. Current modeling and prediction protocols often address uncertainties in an indirect way by reporting several models ranked under some metric with the hope that at least one of the predicted models is close to the truth. However, it is not clear how to ascertain the quality or confidence on individual models of the list, and also there is no statistical guarantee that a near-accurate model is present in the entire list. Protocols for computing specific properties of molecules like surface area, interface area, binding free energy, etc. sometimes provide theoretical guarantees on the approximation errors due to computational approximations, discretization etc., but does not address the inherent uncertainty of the input itself.

Careful practitioners understand that even high quality models derived from x-ray crystallography have uncertainties in the reported atomic coordinates, expressed using B-factors or temperature factors which is correlated to the standard deviation of the reported coordinate; models from electron microscopy also has reconstruction uncertainties attached to each voxel of a 3D scalar field (Fig \ref{fig:pb-uq}A). Current visualization and modeling tools (for example, PyMol \cite{PyMOL}, Chimera \cite{Chimera}, Coot \cite{coot}, JMol \cite{jmol} etc.) allow one to visualize these uncertainties using color maps on the atoms, smooth surface and the scalar field (or volume). However, they fail to highlight the functional relevance of these uncertainties. For example, if the QOI is the optimal conformation of a ligand when its binds to a particular protein found using computational means (typically by finding the minima of a scoring function (Fig \ref{fig:pb-uq}C)), the uncertainties of the atoms near the binding site would have a higher effect on the QOI. We developed visualization techniques that reflect exactly how the uncertainties of different atoms affect the QOI. See Figure \ref{fig:pb-uq}B for example. This new visualization provides functionally important information about the molecule and helps the users direct their focus to more significant sets of atoms, and thereby guiding any rational design effort.

Visualization has to be tied to specific QOIs and their interpretation and relevance to the users. In this article we present several relevant, quantitative and easy to interpret new visualization techniques and enhance some existing ones to aid each step of a typical molecular modeling pipeline. 

\begin{figure*}[h!]
 \centering
\includegraphics[width=0.85\linewidth]{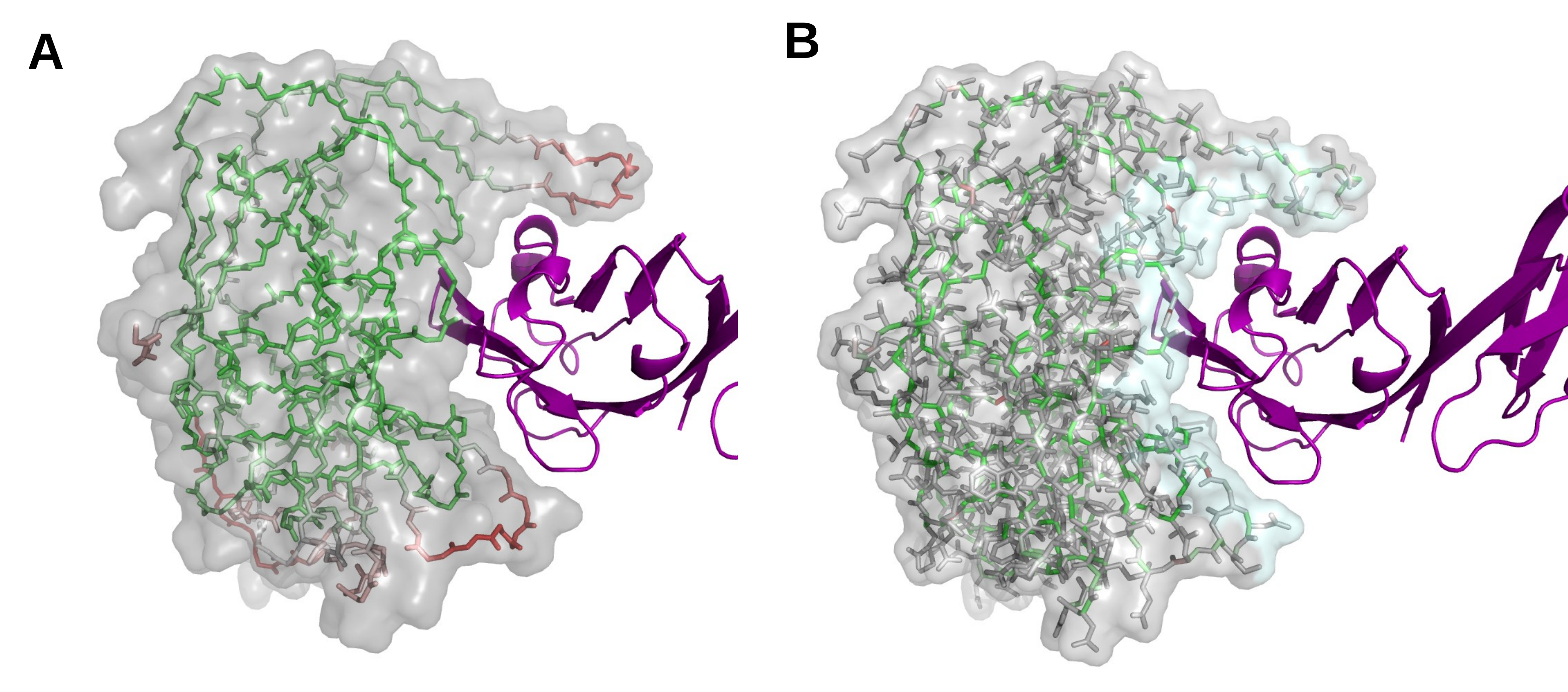}
\caption{\textbf{Visualization of uncertainty of model parameters.} (A) Shows a visualization of the magnitude of variability/uncertainty expressed as B-factors in x-ray crystallography models. Green-white-red color gradient is mapped to atoms with low-medium-high B-factors. This visualization is currently available in PyMol, JMol etc. (B) shows a visualization using the same color-mapping idea, but highlighting the uncertainty in the internal coordinates. In particular edges which have a highly variable dihedral angle (observed across a set of structures of the same molecule) are colored red. This visualization, currently not available in any software, is highly instructive for a molecular modeller since it allows him/her to directly identify key internal angles that need to be carefully sampled. Note that there are fewer internal coordinates which are highly variable and hence allows one to restrict the number of degrees of freedom.}
\label{fig:torsions}
\end{figure*}

At the first step, when one is trying to ascertain the quality and correctness of existing structures, current tools only allows one to visualize the uncertainties in the XYZ coordinate values of a single existing model (Fig \ref{fig:pb-uq}A). We have developed visualizations which summarizes uncertainties from multiple structures/models of the same protein or fragment available in the PDB, to highlight not only the magnitude of uncertainties in Cartesian coordinates, but also uncertainties in the torsion angles for internal coordinates (see Methods) as well as expected principle directions and magnitudes of motion for the atoms. For protein complexes, we also highlight the parts of the structure which belongs to the interaction site. These visualizations immediately expose the uncertainties in the parameter space and also allows the user to choose a subset of parameters to sample during optimization (see Figures \ref{fig:torsions} and \ref{fig:modes}). 

In the next stage, given the available degrees of freedom, one needs to sample and explore the configurational space $\mathcal{C}$. We have developed an extension of pseudo-random number generators for high dimensional spaces which guarantees low discrepancy sampling using a set $\mathcal{C}_N$ of $N$ samples from $\mathcal{C}$ where $N = O(d^c)$, where $d$ is the number of degrees of freedom and $c$ is a constant (see Appendix). By ensuring low discrepancy, and by definition low dispersion, we guarantee that at least one sample $s \in \mathcal{C}_N$ is close to the goal structure $s^*$. The sampling comes with quantification of the dispersion, as well as two separate visualization techniques to visually verify the coverage of the samples (see Figure \ref{fig:sampling}).

The same pseudo-random sampling idea is also applied to bound the uncertainty, arising from uncertain input coordinates, or uncertainties from the previous stage of modeling, in computed scoring functions $\mathcal{F}$. For instance, given a single configuration $s \in \mathcal{C}$, the parameterization of $s$ may include uncertainties which should be propagated to the calculation of $\mathcal{F}(s)$ (see Figure \ref{fig:torsions}). Additionally, we developed quantification visualization techniques to highlight the sensitivity of $\mathcal{F}$ localized to specific atoms/regions of $s$ (see Figures \ref{fig:surfaceuq} and \ref{fig:energyuq}), which aids in adaptive sampling of the space.

Finally, once the user is satisfied with the quality and robustness of the calculation of $\mathcal{F}$, as well as the exploration of $\mathcal{C}$, s/he can confidently rely on the computational protocol to derive meaningful biological insights. In this article, we present three different use cases relevant to drug design along with accompanying visualizations. In drug design, one is typically interested in finding the best drug that would inhibit an unwanted protein-protein interaction, by binding at the same site (and blocking it). We show how one can apply probabilistic binding site analysis to screen multiple drug candidates and identify the best lead (see Figure \ref{fig:bindingsite1}). Given specific targets, and drug candidates (ligands), one typically want to optimize their specificity and binding free energy. We show how a similar probabilistic binding site analysis can lead to identifying the conformation of the target which shows the highest specificity of binding at the expected site (see Figure \ref{fig:bindingsite2}). Finally, we developed visual queues which highlight the relative qualities of multiple ligand conformations (see Figure \ref{fig:bindingsite3}).


Along with identification of important features and attributes whose uncertainty need to be visualized as well as designing intuitive modes of visualization, an important challenge is to rigorously and efficiently compute provable uncertainty bounds in the first place. Much of the prior work on identifying and fixing errors that arise from electron cloud disparity has been focused on solving the problem of distinguishing multiple conformations. A very thorough (although still not exhaustive) listing of methods providing solutions to this problem has been done by Touw {\it et. al} \cite{Touw2014}, but these methods still seek to identify additional information that can be gleaned from the uncertainty (such as identifying information beyond the traditional electron density cloud cutoff \cite{lang2010}, etc.), with very few providing certificates or guarantees on the remaining error.

In this article we pose the problem of uncertainty quantification in a statistical framework, identify key sources of uncertainty in molecular modeling, present theoretical bounds on the uncertainties on the computational outcome, and provide a protocol for bounding the effect of uncertainty/error propagated from one stage of calculations to another. Specifically, for each prediction $f$ of a quantity of interest (QOI), we provide a certificate of accuracy in the form of a Chernoff-like tail bound, i.e. $Pr[|f -E[f]| > t] < \epsilon$, where $E[f]$ is the expectation of the true value and $t$ and $\epsilon$ are two constants. For several biophysical quantities of interest, the underlying random variables need not be independent owing to correlations. In such cases, we relax the assumption on the random variables to being martingales. The correct analogue of the Chernoff-Hoeffding bound is the Azuma inequality. Most stochastic processes can be formulated as a Doob martingale and the Azuma inequality applied to such a martingale becomes what is known as the McDiarmid's inequality. In the Appendix of this article we show how such bounds can be computed theoretically for functions expressed as summation of pairwise distance dependent kernels (e.g. the van der Waals potential). 

Theoretical tail bounds under the McDiarmid model often tend to be too conservative (capturing the worst possible case). However, in practice, empirical analysis using Monte Carlo sampling of the uncertainty space can lead to an approximation of the distribution of the values for the QOI, whose expectation can be used to estimate the tail bounds. This framework was applied to empirically estimate the uncertainties in the computation of surface area (SA), volume, internal van der Waals energy or Lennard-Jones potential (LJ), coulombic energy (CE), and solvation energy under both generalized Born (GB) and Possion-Boltman (PB) models for single molecules; as well as interface area, and binding free energy calculation for pairs of bound molecules. See Appendix for details on the energy functions.

We developed scripts and tools which implement the mathematical framework of sampling, use the existing tools to compute the QOIs \cite{Amber,F2DockGBRerank,BajajChenRand11,modeller,AutoDock}, compute uncertainty bounds as well the visualization directives which can be directly loaded into existing molecular, surface and volume visualization software \cite{TexMol2004,PyMOL}.  Our methods should enable the end user of these tools achieve a more quantitative and visual evaluation of various molecular models for structural and property correctness, or the lack thereof.

\hide{
Under our statistical framework, we use these B-factors and define the position of each atom as a random variable (instead of a fixed coordinate) which is distributed normally (Gaussian). Hence the structure of the entire molecule is expressed as a joint distribution of independent normal distributions. Assuming that $\mathcal{X}$ denotes the space of possible structures represented by the joint distribution of B-factors for a given x-ray structure $X$, we can provide the following set of uncertainty measures for a particular protocol $P$ that computes a quantity $f(X)$ for $X$ (without accounting for the uncertainties in $X$)-

\begin{itemize}
 \item A certificate for $P$: The probability, that $f(X)$ computed using only the original model $X$ does not have an absolute error more than some constant $t$, is less than a user-defined acceptance threshold $\epsilon$ (i.e.\ $Pr[|f(X) -E[f]| > t] < \epsilon$ where $E[f]$ is the expected value of $f$ over $\mathcal{X}$).
 \item A certificate for the structure/model $X$: The probability that any perturbation of the atoms under the joint distribution based on B-factors would lead to a deviation of $f$ beyond a particular threshold $t$ is less than some constant $\epsilon$ (i.e.\ $\forall_{X' \in \mathcal{X}} Pr[|f(X') - E[f]| > t] < \epsilon$.
\end{itemize}
}

\section{Methods}
\subsection{Parameterizations and uncertainties of molecular models}
Molecular structural models are typically parameterized using either a list of their XYZ coordinates, or using internal coordinates (which is a series of bond length, bond angle and dihedral angle). In the first representation the degrees of freedom or the space of configurational uncertainty is related to each coordinate value; in the latter representation, typically the dihedral angles are the only degrees of freedom since bond lengths and angles are considered constants. 

x-ray crystallography experiments reconstruct a 3D electron density cloud from the diffraction pattern generated from a crystal lattice of the molecule. For high resolution reconstructed electron densities, expected locations of individual atoms can be identified. Hence, it is common to report such models using the XYZ coordinates of the atoms. However, expected location is not necessarily perfectly determined. There is a degree of uncertainty which is expressed as temperature factors or B-factors. Simply stated B-factors are a measure of the error in the match and fit of specific atoms within the electron density cloud constrained by the protein's primary, secondary and tertiary structure and inter-atom biochemical/biophysical forces. 

B-factors are derived from structure factors, which are based on the Fourier transform of the average density of the scattering matter. The structure factor, $F(\vec{h})$, for a given reflection vector, $\vec{h}$, is the sum of the optimized parameters for each atom type $j$, and position $\vec{x_j}$ and as defined by the following equation:

$$F(\vec{h}) = \sum_j f_j \exp\left(-\frac{1}{4} B_j \vec{h}^{\,t}\vec{h}\right) \exp\left(2\pi i \vec{h}^{\,t}\vec{x_j}\right),$$
where $f_j$ is the scattering factor, $B_j$ is the B-factor for atom $j$, and $\vec{x_j}$ is the 3-dimensional position of each atom \cite{schneider1996}.

If we assume that the static atomic electron densities have spherical symmetry (or, more specifically defined by a trivariate Gaussian, $\vec{u}$), this can be converted into the anisotropic temperature factor commonly used, $T(\vec{h})$ \cite{trueblood1996}:

$$T(\vec{h}) = \exp\left[-2\pi^2\langle(\vec{h}\cdot\vec{u})^2\rangle\right],$$
where the univariate Gaussian form (needing not the direction of $\vec{h}$, but only its magnitude) is described by:

\begin{equation}
T(|\vec{h}|) = \exp\left[-8\pi^2 \langle u^2\rangle (\sin^2\theta)/\lambda^2\right].
\label{eq:bfactor}
\end{equation}

Finally, the B-factor is defined as $B=8\pi \langle u^2\rangle$. Thus, a B-factor of 20, 80, or 180\AA$^2$ corresponds to a mean positional displacement error  of 0.5, 1, and 1.5\AA, respectively. 

The internal coordinate system, the assumption is that the entire molecule is like a connected graph embedded in 3D space. Each node of the graph is an atom, and each edge represent a bond. So, given the position of any one atom, and the bond length, bond angle (angle at a node between two bonds), and dihedral angles (given three successive bonds, the dihedral angle is defined as the angle between the two planes formed by bonds 1,2 and bonds 2,3); the location of all atoms can be uniquely determined. Moreover, internal coordinates successfully captures the dependence of atom positions on the positions of the neighbors. Also, since bond lengths and angles have been empirically observed to be constants, this representation allows one to reduce the number of degrees of freedom to only the change of dihedral angles (henceforth called torsion angles since the change is similar to a twisting motion around a bond).

\begin{figure*}[ht!]
 \centering
\includegraphics[width=0.85\linewidth]{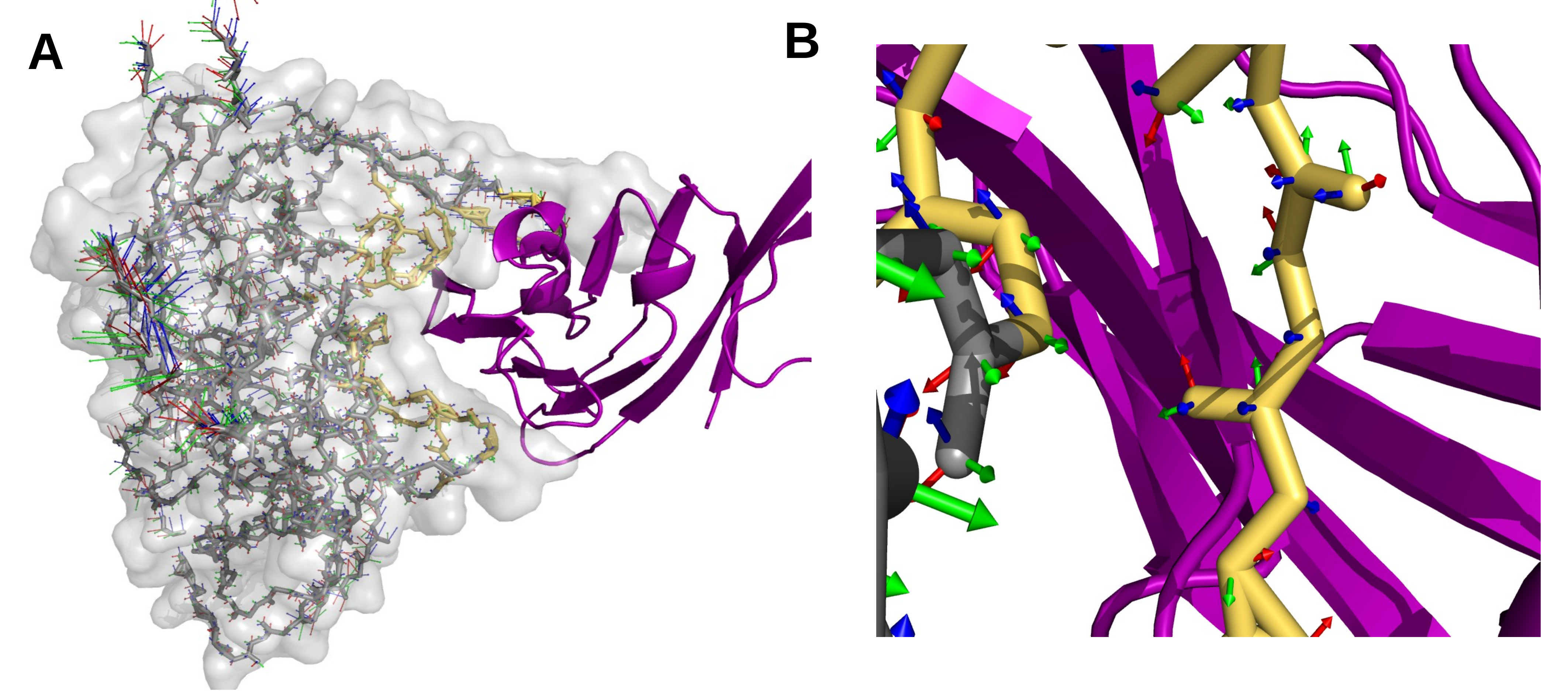}
\caption{\textbf{Visualization of uncertainty of atomic positions given an ensemble of models.} (A) Shows the bonded backbone structure of protein gp120, in complex with antibody CD4 (ribbon model in purple). The visualization summarizes the location data for each atom of the backbone from all available structural models (of gp120) from the protein data bank. Three vectors at each atom shows the variability of the location of the corresponding atom, as well the primary directions of its movement. The length of the vectors represent the magnitude of the variance of the corresponding atom location. The vectors are colored red/green/blue simply for the sake of distinction and does not carry any quantitative meaning. The part of the backbone which is at the interface with CD4 is highlighted in yellow. This visualization clearly shows that while parts of gp120 are highly variable, most of them are not at the interface with CD4. A zoomed in view of the interface region is shown in (B).}
\label{fig:modes}
\end{figure*}

\subsection{Statistical framework for uncertainty quantification and propagation}
One typically defines statistical uncertainty quantification as a tail bound, namely, a probabilistic certificate as a function of  a parameter $t$, where the computed value $f(\mathbf{X})$ of a QOI expressed as some complicated function or optimization functional involving {\sf noisy} data $\mathbf{X}$ is not more that $t$ away from the true value, with high probability. Or alternatively the probability of the error being greater than or equal to $t$ is very small. Such a certificate is expressed as a Chernoff-Hoeffding \cite{Chernoff_1952} like bound as follows-

\begin{equation}
Prob (f,X, t,\epsilon) = \Pr[|f(X) - E[f]| > t] \leq \epsilon.\
\label{eqn:chern-x}
\end{equation}

For the special case when $Y$ is a linear combination of the components of $\mathbf{X}$, simple statistical attributes like variance, standard deviation etc. of $Y$ can be computed directly from the variances of $\mathbf{X}$, even if the components are not independent. If, $Y$ is not a linear combination, then an interval approximation (confidence intervals, likelihoods etc.) can be computed using a Taylor-like expansion of the moments of $Y$ \cite{Goodman_1960}.

In this article, we adopt a method of bounded differences, which is a modification of Markov and Chebyshev inequalities, used for independent random variables to derive Chernoff-like bounds. We briefly discuss the derivation of a loose uncertainty bound based on Doob martingales as introduced by Azuma \cite{Azuma_1967} and Hoeffding \cite{Hoeffding_1963} and later extended by McDiarmid \cite{McDiarmid_1989}. Then we proceed to apply this model to bound the uncertainty in molecular modeling tasks that involve summations over decaying kernels (e.g. electrostatic interactions). We also extend the applicability of McDiarmid-like bounds to cases where the random variables are not necessarily independent (see Appendix for details).

Since theoretical upper bounds often overestimate the error, we also explore an alternate quasi Monte Carlo (QMC) approach\cite{Niederreiter_1990,James_Hoogland_Kleiss_1996}. The QMC is useful in approximating the distribution of values for the QOI under a given statistical model of the input uncertainty, and then empirically establish the uncertainty of individual values of the QOI, as well as providing certificates like Equation \ref{eqn:chern-x}. Although the quality of the QMC based certificate itself depends on the quality and size of the samples, our experiments (see Results and Discussion) showed that typically fewer than 200 samples under a low discrepancy sampling techniques provides sufficiently accurate approximations of the certificates.

\subsubsection{Empirical uncertainty quantification using QMC}
Quasi Monte Carlo method of uncertainty propagation, it is assumed that $X_i$ are independent random variables and their probability distribution functions (PDFs) are known. Now, a low discrepancy sampling of the product space $X_1 \times X_2 \times \ldots \times X_n$ must be generated which would define an approximate PDF for $Y$ from which we could compute all the  necessary tail bounds. We explore several such product spaces and use low discrepancy  sampling strategies to derive PDF's of functions $Y$ with bounded error and guaranteed convergence. Note that a simpler Quasi Monte Carlo (QMC) sampling can be applied to find the minimum $c_k$ for each $X_k$, and hence derive the overall bound. Hence the most crucial component of UQ and UP under this QMC framework boils down to identification of the set of independent random variables $\mathbf{X}$, which affect the computation of the QOI ($Y$), along with their approximate PDFs and corresponding sampling techniques.

\subsection{Defining and sampling the space of configurations}
The statistical framework described above, requires geometric models of the molecules, parameterization of the degrees of freedom available to the molecule, a mapping which allows one to update the model based on sampled degrees of freedom, and an implementation of the QOI. This this section, we describe how we derive the relevant degrees of freedom and set up the joint probability distribution which gets sampled by the QMC protocol. We explore both vibrational/positional uncertainty variables (from B-factors), and flexible/configutational uncertainty variables (internal torsion angles).

Given the x-ray structure $M$ containing $n$ atoms of a protein or a complex of two proteins in the PDB file format, we extract the anisotropic B-factors $B_i^x, B_i^y$ and $B_i^z$ for each atom $a_i \in M$. The distribution of the position of the atom in each direction is modeled as a Gaussian distribution whose PDF is defined as $p(x_i)=\frac{1}{\sigma_i^x\sqrt{2\pi}} \exp^{-\frac{(x_i-\mu_i^x)^2}{2\sigma^2}}$ where $\sigma_i^x$ is the standard deviation derived as $\sqrt{\frac{B_i^x}{8\pi}}$ from the B-factor, and the mean $\mu_i^x$ is the expected position of the atom. Note that for some x-ray structures only an isotropic B-factor $B_i$ is reported. In that case we simply assume $B_i^x = B_i^y = B_i^z = B_i$. As mentioned earlier, we consider these distributions to be independent consistent with the assumptions made while computing the B-factors themselves.

\begin{figure}[ht!]
 \centering
\includegraphics[width=0.95\linewidth]{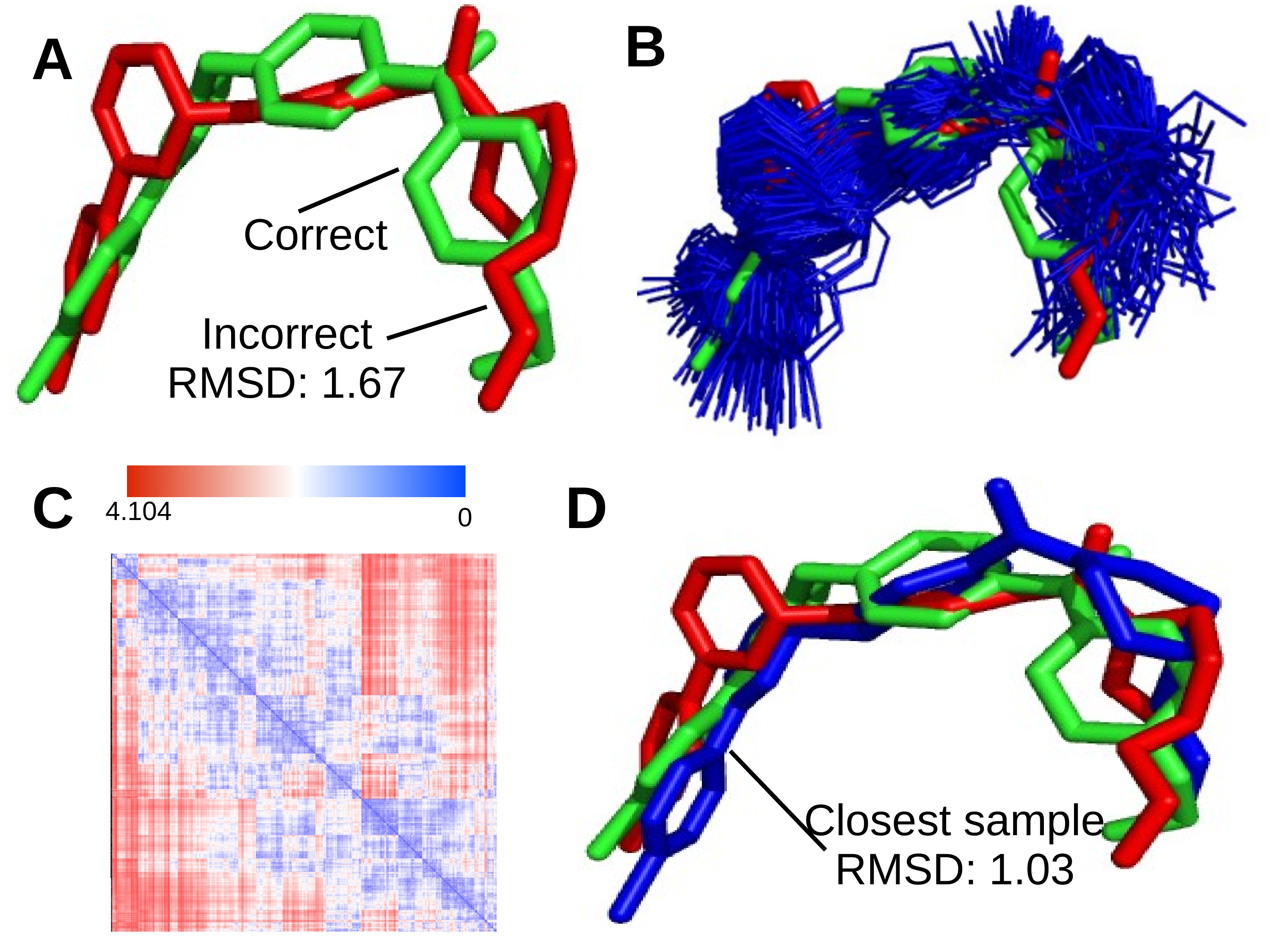}
\caption{\textbf{Sampling the space of configurations.} In this figure we explore the application of sampling to find the correct conformation of a ligand. (A) Shows two conformations of the same ligand, available in two different models in the protein data bank. The red one was deposited earlier, was found to be incorrect later, and was replaced by the one shown in green. (B) We show that applying the pseudo-random sampling on the space defined by the internal coordinates of the bad ligand model allows one to generate an ensemble of structures (shown in blue). (C) provides a heatmap of the distance (in terms of RMSD) between the samples which shows the randomness of the set of samples. (D) shows one of the samples (blue) along with the good and bad conformations and shows that the sampling includes a model close to the good one.}
\label{fig:sampling}
\end{figure} 

For the internal coordinates, we have already mentioned that the bond lengths and angles are considered constants. The dihedral angles are also constant for some specific motifs, like an aromatic ring, or the peptide plane. However, there are still some dihedral angles which are free. For proteins, these consists of two angles on each peptide (a protein is a chain of peptides) defined as $\phi$ and $\psi$, as well as zero or more free dihedrals on the side chain of each peptide. For the purpose of this article, we shall simply assume that a total of $m$ such dihedral anlges are free. We shall express each such degree of freedom using a random variable distributed uniformly between a range $[lower,upper]$ where the limits of the range is derived from the so called Ramachandran plot \cite{Ramachandran}, which is an empirical study of dihedral angle values observed in protein structures. Note that, the dihedral angle degrees of freedom are truly independent.

\subsubsection{Sampling}
The joint distribution, either defined as the product space of the $3n$ independent Gaussian distributions for the B-factor case, or the product space of the $m$ independent uniform distributions for the torsion angles with within lower and upper limits defined by Ramachandran plots, represents the space of possible configurations for the molecule. Hence, sample the joint distribution, we first sample the space $[0,1]^{3n}$ using pseudorandom generator which guarantees low discrepancy sampling in high dimensional product spaces \cite{GMRZ_13} to produce a tuple $<u_1^x, \ldots, u_n^z>$. For the B-factor case, each number $u_i^j$ from this tuple is mapped to get a sample $a_i^j$ from a Normal distribution using the Box-Muller method \cite{box1958note}, and finally appropriate translation and scaling is performed to get a sample from the corresponding Gaussian distribution as follows $s_i^j = \mu_i^j + \sigma_i^ja_i^j$. These samples $<s_1^x, \ldots, s_n^z>$ are used to displace the atoms to generate a new configuration. For the torsion angle case, the $u_i^j$ is mapped from $[0,1]$ to the $[lower,upper]$ range by simple uniform scaling, and then the resulting torsion angles are applied using the Denavit-Hartenberg transformation \cite{Hartenberg_Denavit_1955} to produce a new geometric model of the molecule. Figure \ref{fig:sampling} shows an application of the internal coordinates (dihedral angle) sampling applied to generate 1000 samples from a poorly configured model of an inhibitory ligand for kB kinase $\beta$ (PDBID:3QAD). A well configured model for the same ligand is already available (PDBID:2RZF). In Figure \ref{fig:sampling}(A) we have contrasted the bad and good configurations. The RMSD between the models is 4.667. Figure \ref{fig:sampling}(B) shows the 1000 samples we generated. A low dispersion sampling guarantees that at least one sample would be close to the good configuration. Indeed, in Figure \ref{fig:sampling}(D) we show one of the several samples which had low RMSD compared to the good configuration.

An important point to note is that the above procedure, while maintaining the constraints implied by the B-factors and the mean positions and only making small perturbations, does not necessarily guarantee that the new configurations will be biophysically feasible. In particular they do not preserve the bond angle and bond lengths. So, we use the Amber force field \cite{cornell1995second} and energy minimization with implicit solvents to relax the sampled structures. We rejected any sample for which the relaxed structure still had high van der Waals potential indicating that some atoms are too close to each other. Accepted models were protonated and assigned partial charges based on the Amber force field using the PDB2PQR program \cite{Dolinsky04}.

\begin{figure}[h]
 \centerline{\includegraphics[width=4in]{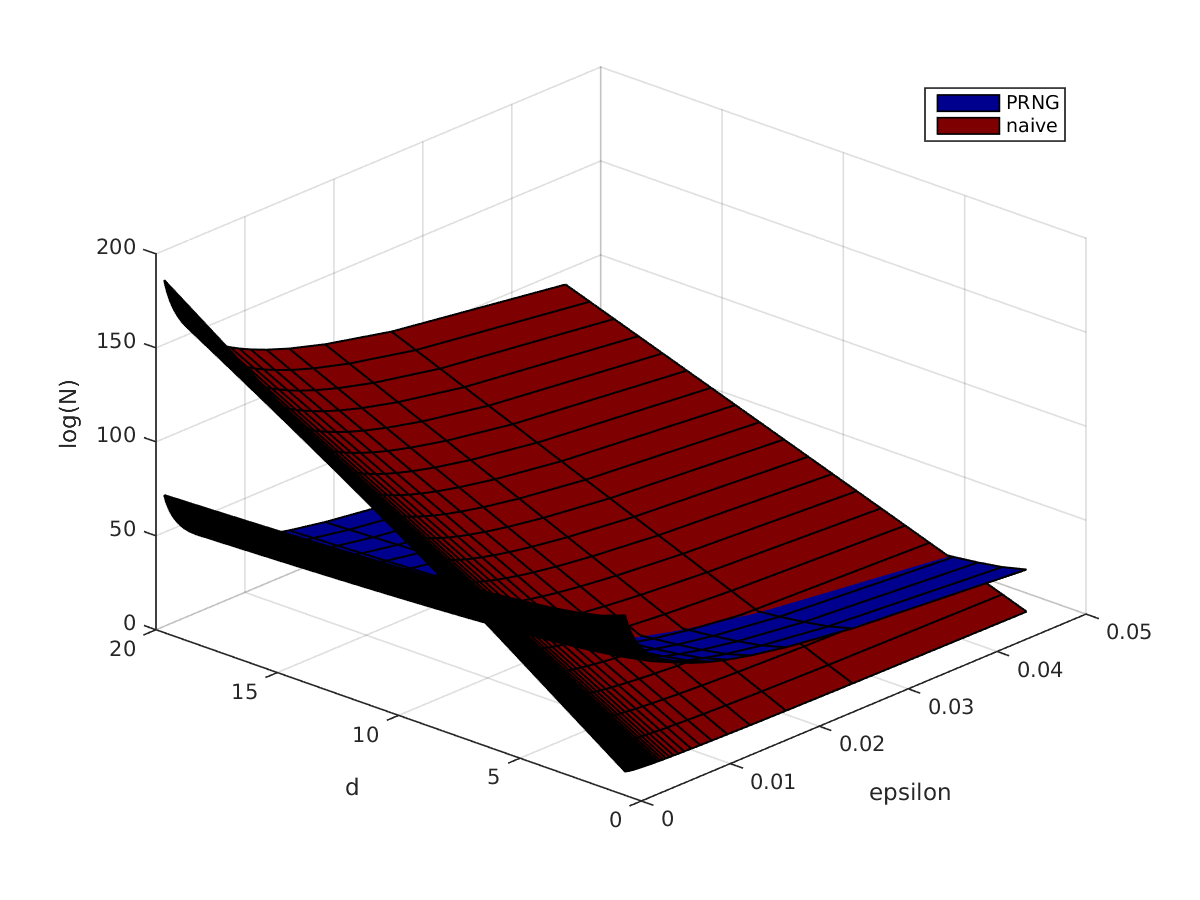}}
 \caption{Number of samples needed to maintain $\epsilon$ discrepancy. Note that the y-axis is log-scale, so the number of samples ($N$) required for the naive method grows exponentially in $d$ (number of dimensions), whereas the pseudorandom method (PRNG) grows as a function of $\log{d}$.}\label{fig:prng_vs_native}
\end{figure}

One of the major drawbacks with QMC sampling is that the curse of dimensionality prevents naive sampling with a large number of dimensions. Assuming that one wants $\epsilon$ discrepancy for each random variable, which requires $m$ samples in a single dimension (assuming $d$ r.v.'s, whether from the $3n$ independent Gaussians or $\nu$ torsion angles). Then to maintain the same level of discrepancy, the QMC protocol would require $m^d$ samples. For any reasonable measure of discrepancy (i.e. $<1\%$), this quickly becomes intractable. 

For this reason, it is absolutely crucial to use a sampling method that requires a polynomial number of samples. The low-discrepancy product-space sampler developed by Bajaj {\it et al.} \cite{BBCZ_2015} reduces the number of samples significantly from $m^d$ to only $\left(\frac{d}{\epsilon}\right)^{O(\sqrt{\log{\left(\frac{1}{\epsilon}\right)}})}$, where $m=\left(\frac{d}{\epsilon}\right)^{3+o(1)}$. Note that this is polynomial in $m$ and $d$. See Figure~\ref{fig:prng_vs_native} for a summary of the number of samples required for different values of $d$ and $\epsilon$. With this sampling method, we can produce low-discrepancy bounds without a computationally-intractable algorithm.

\subsection{Quantifying and visualizing uncertainties in molecular properties and scoring functions}
While the visualizations in Figures \ref{fig:torsions} and \ref{fig:modes} highlight the inherent uncertainties in the molecular structure and their parameterization, they do not directly highlight the effect of these uncertainties on computed properties of the molecule. Specifically, we are interested in bounding the propagated uncertainty in the calculated property, and also localize the origins of uncertainty which disproportionately affect the outcome. This is carried out using the statistical QMC framework described above. We sample an ensemble of structures based on the inherent uncertainties of the model, compute the quantity of interest, compute certificates of accuracy etc. See Results for a detailed quantitative analysis in terms of bounding the uncertainties in computed area, volume, and solvation energies. Now we discuss some techniques which allows one to visually explore such uncertainties.

\subsubsection{Pseudo-electron cloud}
\begin{figure}[ht!]
 \centering
\includegraphics[width=0.95\linewidth]{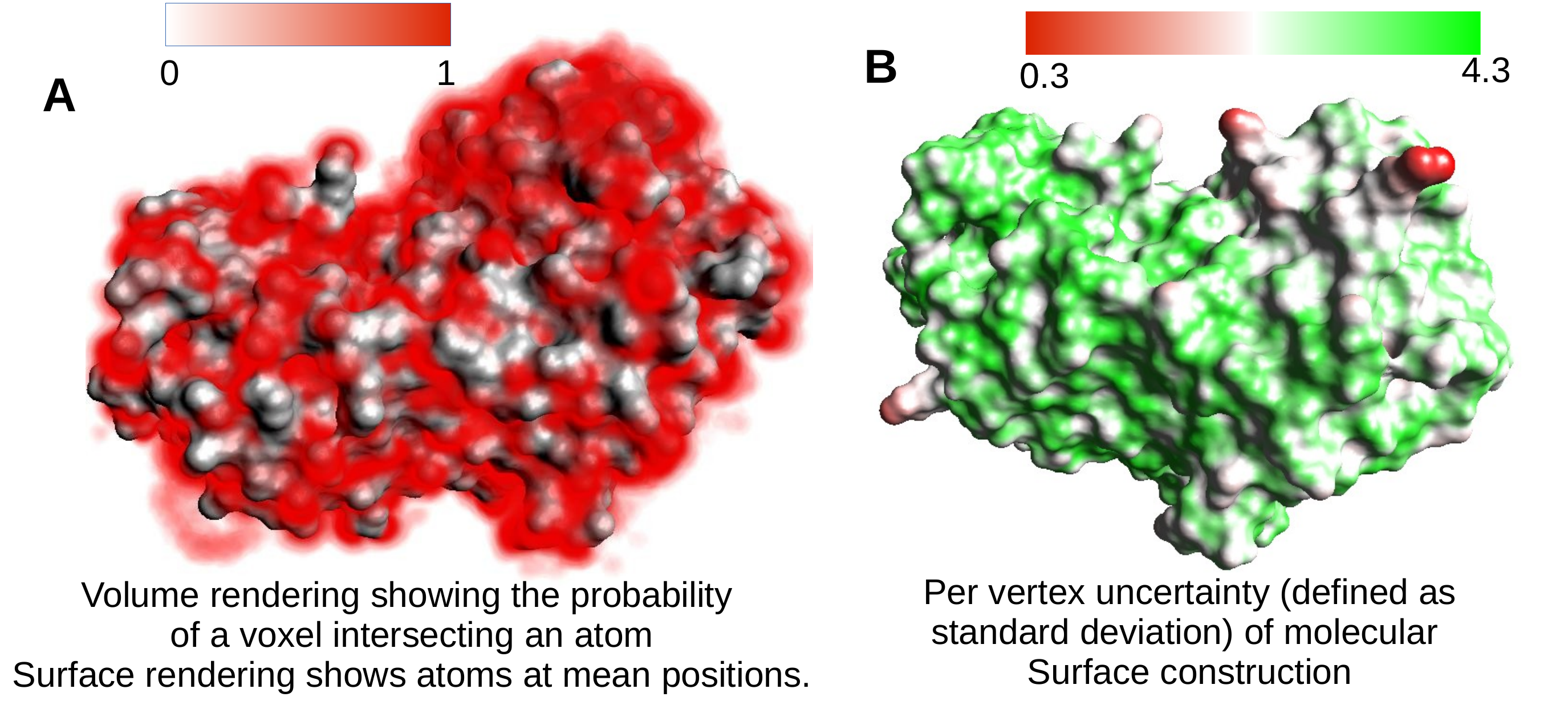}
\caption{\textbf{Visualization of molecular surface uncertainties.} (A) A volumetric map showing the likelihood of the voxel being occupied by an atom, computed using a sampling of the joint probability distribution of the atom positions. (B) Expected deviation of each point on the surface of a single model, w.r.t.\ all models sampled based on the joint distribution of the locations of the atoms. Green colored regions are expected to remain more or less stable in any sample, red colored ones may vary a lot.}
\label{fig:surfaceuq}
\end{figure}

We propose a visualization where the samples are combined into a single volumetric map whose voxels represent the likelihood (over the set of samples) of an atom occupying the voxel. Figure \ref{fig:surfaceuq}(A) shows such a visualization for 1OPH:A. Note that this data is not simply useful for visualization, but can be used as the representation of the shape of the molecule for docking and fitting exercises to incorporate the input uncertainties directly into the scoring functions.

\subsubsection{Localized uncertainty/error in molecular surface calculations}
In many applications, instead of a volumetric map, one uses a smooth surface model to compute QOIs like area, volume, curvature, interface area etc. In such cases, a visualization like Figure \ref{fig:surfaceuq}(B) can be very descriptive. It shows a single smooth surface model (based on the original/mean coordinates), and the colors at each point on the surface show the average distance of that point from all surfaces generated by sampling the joint distribution. Unsurprisingly, most parts of the surface in the figure has very low deviation, and only the narrow and dangling parts have high deviation. Comparing this with the rendering of B-factors (in Figure \ref{fig:pb-uq}(A)) shows that even though some parts of the surface are in regions with high B-factors, the uncertainties does not affect the surface computation as much. 

\begin{figure}[ht!]
 \centering
\includegraphics[width=0.95\linewidth]{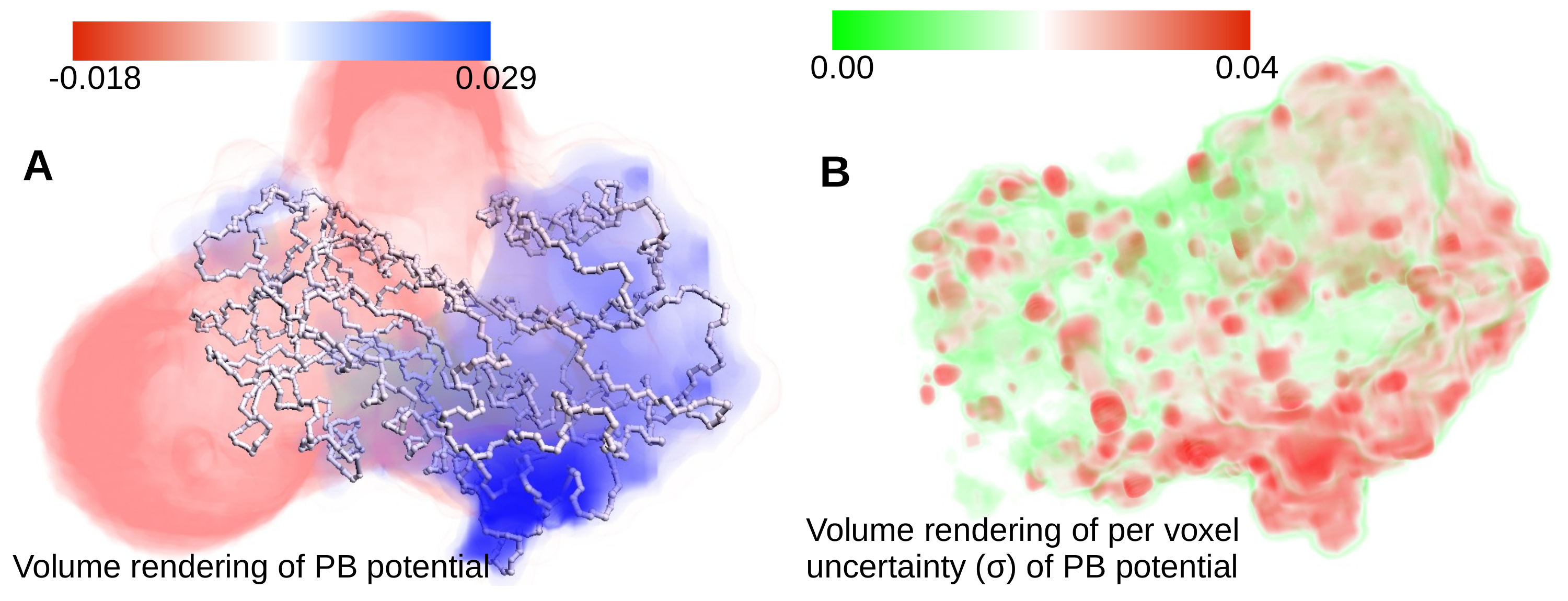}
\caption{\textbf{Visualization of molecular surface uncertainties.} (A) A volumetric map showing the likelihood of the voxel being occupied by an atom, computed using a sampling of the joint probability distribution of the atom positions. (B) Expected deviation of each point on the surface of a single model, w.r.t.\ all models sampled based on the joint distribution of the locations of the atoms. Green colored regions are expected to remain more or less stable in any sample, red colored ones may vary a lot.}
\label{fig:energyuq}
\end{figure} 

\subsubsection{Uncertainty in energy calculations}
Now we focus on QOIs that are computed based on other intermediate QOIs. For example, computation of PB energy first evaluates the PB potential on a volume which encapsulates the molecule and the solvent. This potential calculation itself requires a smooth surface representation of the molecule as input, along with the positions and charges of the atoms. In this case, as well as bounding the overall uncertainty for the final value of the PB energy, we can also bound the uncertainties of the intermediate PB potentials calculated at each Voxel. We do this by defining the PB potential at a voxel as separate QOIs and apply the QMC sampling to generate an ensemble of atomic models and smooth surfaces, then evaluate the QOIs for each sample. Hence, we derive a distribution for each voxel. The means of these distributions are rendered in Figure \ref{fig:energyuq}(A) showing the negative and positive potential regions. The standard deviations of the distributions are rendered in Figure \ref{fig:energyuq}(B). Comparing the original uncertainties (B-factors in Figure \ref{fig:pb-uq}(A)) to these third level propagated uncertainties shows that while in some regions the uncertainties had a cancellation effect, in some other regions they amplified.

\subsection{Quantifying and visualizing uncertainties in binding sites}
Computationally predicting the correct binding site on a given protein for binding with a specific ligand or a class of ligands is an important problem in drug design. Often the goal is the inhibit or prevent a protein A from binding to a partner molecule B. So, one tries to first identify the site on A where B is supposed to bind. Then, one would explore a set of drug candidates and try to find one (let us call it C) that would bind to the same site with stronger affinity and thereby prevent B from binding there. In some cases, the configurations of both the protein A and the ligand C must be re-resolved in each others presence, to get a more realistic estimate of the binding strength. 

Typically, protein-protein and protein-ligand binding interactions are computationally predicted by sampling the space of relative orientations (as well as each one's internal configuration) to find the optimal orientation. This procedure is often referred to as docking. However, due to the combined uncertainties of the models, the energy/scoring function calculations, and sampling, one cannot simply assume that the highest scoring model out of the docking procedure. Here, we show that even if one considers the top $k$ results from the docking, then they can be used to define a probabilistic model of the binding site. This model is robust from spurious results and minor perturbations/errors in the scoring.

\subsubsection{Probabilistic model of binding site inferred from computational sampling and scoring}
Given a validated scoring function $\mathcal{F}$ with bounded errors, specific configurations $s_A$ and $s_B$ of two molecules $A$ and $B$, and a low dispersion sampling of the space of relative orientations $SE(3)$, we compute a list of the top $k$ ranked orientations. Let the $i$-th orientation is expressed using a transformation $T_i$ which is applied to $s_B$ (denoted $T_i(s_B)$). Now, for each atom $a$ on molecule $A$, let $BS(a)$ denote a random variable denoting the event that $a$ is in contact with $s_B$ upon binding; i.e.\ $a$ is on the binding site. Now, we define the probability of $BS(a,s_B)$ as follows

\begin{equation}
 p_{BS}(a,s_B) = Prob[BS(a,s_B)] = \frac{1}{k} \sum_i(contact(a,T_i(s_B)))
\end{equation}

\noindent
where $contact(a,T_i(s_B))$ is 1 if at least one atom in $T_i(s_B)$ is within a distance cutoff from $a$, otherwise $contact(a,T_i(s_B))$ is 0.

Hence, we identify the binding site probabilistically. Given an accurate docking tool and molecules in favorable configurations, the probabilities would be high for small contiguous regions of the molecule A, and low in other regions. On the other hand, almost equal probability across the A would indicate poor docking prediction, and/or poor affinity between the molecules. We explore applications of this analysis and visualization in the next few subsections.


\subsubsection{Inhibitor selection based on binding site overlap}
Consider the scenario where the configurations of molecule A, and its binding partner B are known; and we want to select an inhibitor from a set of candidates which would bind to A at the same site as B. Without loss of generality, assume that we want to choose between two candidates $U$ and $V$. We first compute $p_{BS}(a,B)$, $p_{BS}(a,U)$ and $p_{BS}(a,V)$ as described above. Note that we assumed that the binding orientation of A and B is known, hence $p_{BS}(a,B)$ is either 0 or 1. Now, we define the quality of candidate U, in terms of its expected ability to block the binding of B, is expressed as follows-

\begin{equation}
 inhibit(U) = \sum_{a \in A} p_{BS}(a,B) \dot p_{BS}(a,U).
\label{eq:bsiteoverlap}
\end{equation}

\begin{figure}[ht!]
 \centering
\includegraphics[width=0.95\linewidth]{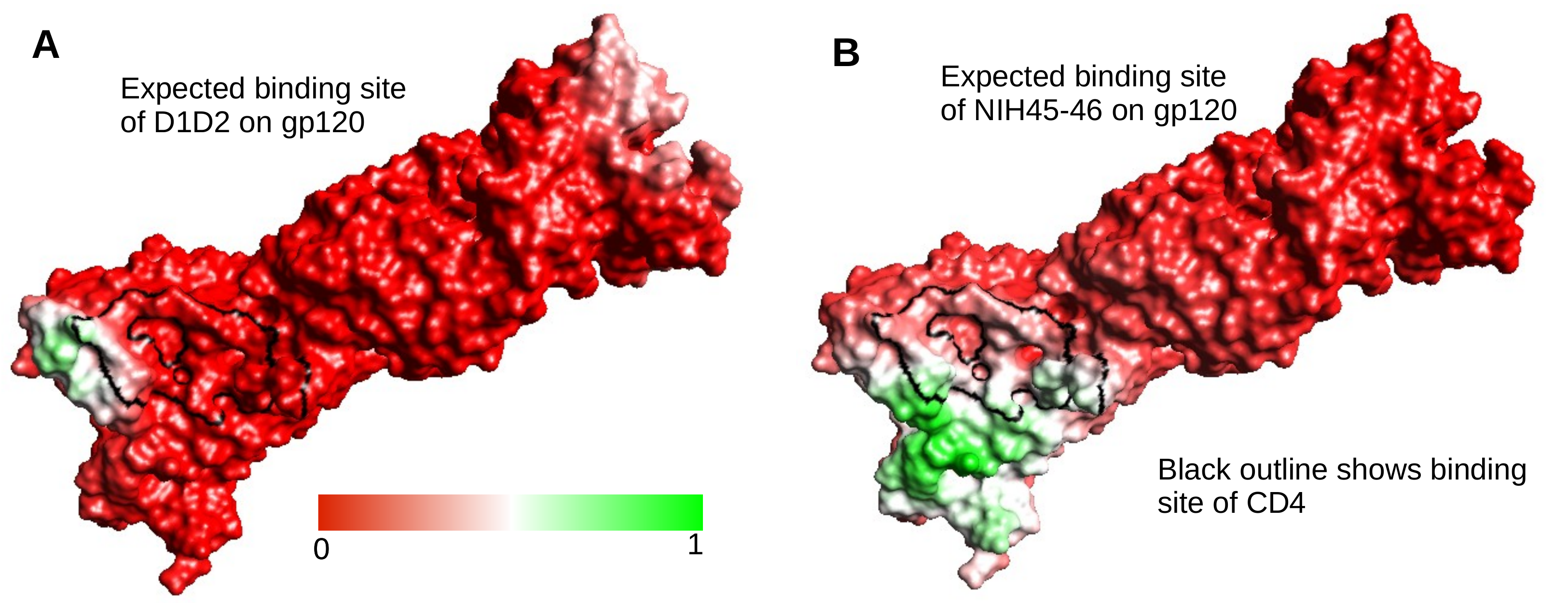}
\caption{\textbf{Comparing binding sites of two different ligands on the same protein.} The probabilistic binding site of D1D2 and NIH45-46, two possible antibodies for inhibiting gp120-CD4 interaction. The binding site of CD4, known from existing models, are highlighted using black outline. The probable binding site of the antibodies, inferred from computational docking, is shown using Green-white-red gradient representing high-medium-low probabilities.}
\label{fig:bindingsite1}
\end{figure}

Comparing $inhibit(U)$ and $inhibitVU)$, one can choose the most likely candidate. We have also developed an intuitive visualization for this scenario, where we delineate the B's binding site on A (boundary of the regions which have $p_{BS}(a,B) = 1$), and then color the surface of A using $p_{BS}(a,U)$. For example, in Figure \ref{fig:bindingsite1}, we explore the efficacy of two antibodies D1D2 and NIH45-46 in preventing the HIV spike protein gp120 from binding with host-cell's membrane protein CD4. It is apparent that NIH45-46 is a better candidate for inhibiting gp120-CD4 binding.

\subsubsection{Model selection based on binding site specificity}
\begin{figure}[ht!]
 \centering
\includegraphics[width=0.95\linewidth]{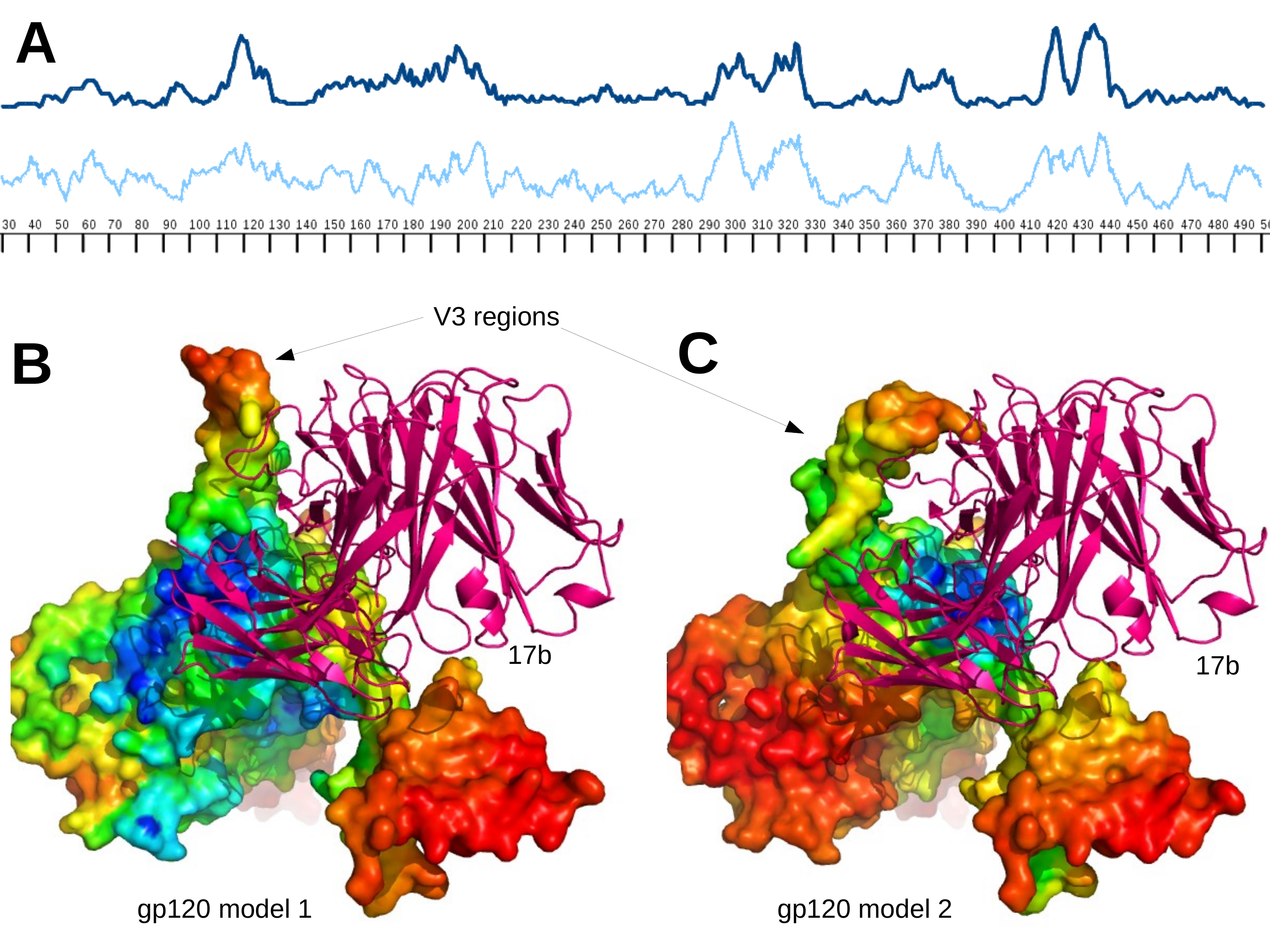}
\caption{\textbf{Comparing binding sites of the same ligand on two different models of the same protein.} Two separate conformations of gp120 was docked with antibody 17b. (A) shows, for each residue along the backbone of gp120, the probability that the residue is on the binding site between that model of gp120 and 17b. (B) and (C) shows the same data, mapped onto the 3D structural models (shown as smooth surfaces) of gp120, using a rainbow colormap where blue represent high probability and red is low probability. A gradient with a higher number of colors is essential in this case to clearly distinguish the binding site specificity of the two models. Note that the model in (B) shows more specificity at the expected binding site (shown by rendering 17b in its expected orientation).}
\label{fig:bindingsite2}
\end{figure}

In many applications the structure of the target is not complete resolved a-priori. For example, the expected orientation of 17b and gp120 is known from low resolution data, but the exact atomic configuration of gp120 near the binding region is not known. In these cases one typically samples the space of possible configurations, select a few which score high according to a scoring function, optimize them and finally determine the best candidate. However, in this case the quality of the structure depends not only on its own internal energy, but also on the affinity and specificity of its binding to the expected ligand. Probabilistic binding site analysis, as described above, can greatly contribute to the selection in this scenario. In Figure \ref{fig:bindingsite2}(A), we show the probability of specific residues (small collections of contiguous atoms) of being on the binding site. Now, we couple this with the prior knowledge of the expected binding site of 17b. A model which has clusters of high probability near the expected binding site definitely offers a better binding interface to 17b, and hence is a better model. This measure is easily expressed mathematically similar to Equation \ref{eq:bsiteoverlap}. 

The visualizations shown in \ref{fig:bindingsite2}(B)-(C) provide immediate confirmation and intuition behind the choice. In this figure, we sampled the configuration of the variable regions of gp120 (which are hard to resolve experimentally), and one of the variable regions, called V3, is especially relevant in binding with 17b. In the figure, we see two different configuration of the V3 regions, and one of them (left panel) seems to favor or allow 17b to bind at the expected site, and the other seems to block the site a little bit and forces 17b to bind slightly off its expected site.

\subsubsection{Model selection based on binding site interactions}
During ligand optimization for binding, one needs to sample the configuration for the ligand and then, for each configurations, apply docking to predict the best ranked orientations (of the sample configuration). In such cases, we augment the definition of the probability of an atom being on the binding site by summing over all configurations, i.e.\

\begin{equation}
 p_{BS}(a,\mathcal{C}^B) = Prob[BS(a,\mathcal{C}^B)] = \frac{1}{kN} \sum_{s_B \in \mathcal{C}^B_N} \sum_i(contact(a,T_i(s_B)))
\label{eq:bindingprob2}
\end{equation}

\noindent
where $\mathcal{C}^B$ represents the configurational space of $B$, and $\mathcal{C}^B_N$ is the $N$ discrete samples taken from this space.

Given this probabilistic estimate of the binding site, we define a new scoring function to rank the ligand configuration+orientations. For a given orientation $T_i$ of a given configuration $s_B$, we define its score as follows-

\begin{equation}
 bindingscore(s_B, T_i) = \sum_{a \in A} p_{BS}(a,\mathcal{C}^B) \dot contact(a,T_i(s_B)).
\label{eq:bindingscore}
\end{equation}

Hence, configurations of the ligand which can bind at highly probably binding sites, are rewarded. 

\begin{figure}[ht!]
 \centering
\includegraphics[width=0.95\linewidth]{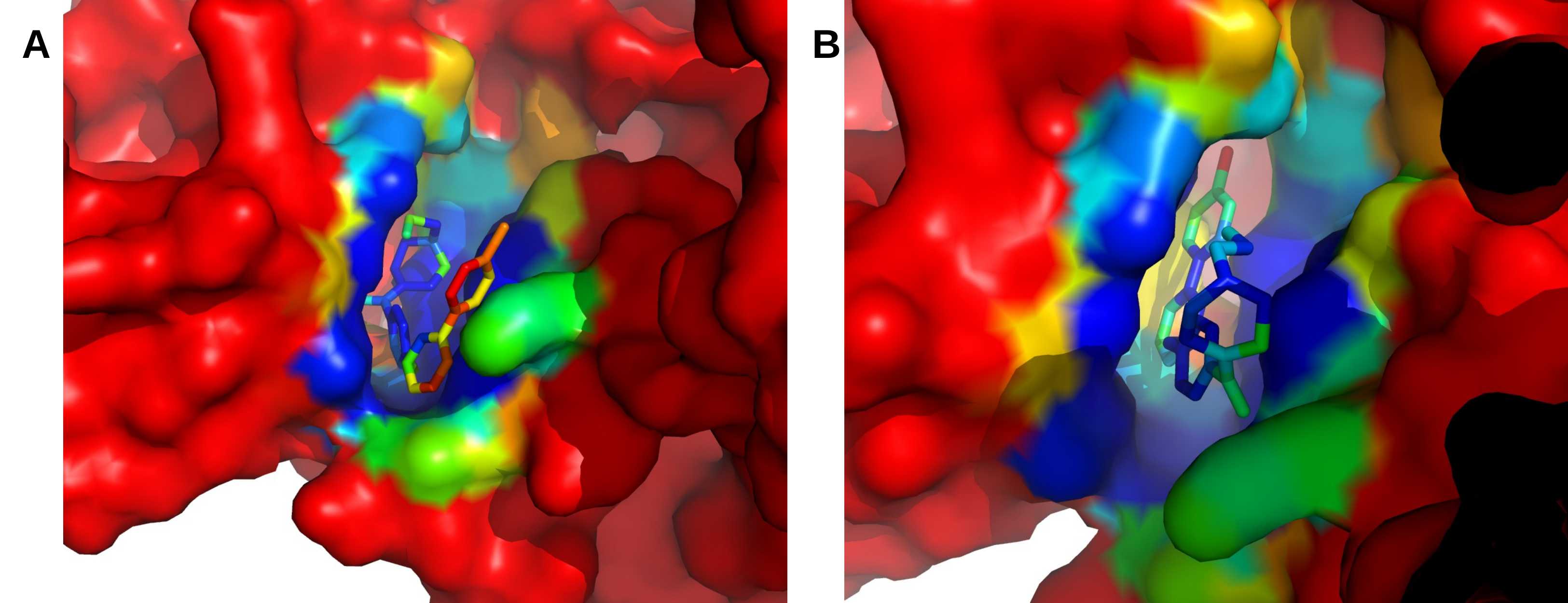}
\caption{\textbf{Comparing different ligand conformations using probabilistic binding site on the same protein.} The smooth surfaces in the figures show the probabilistic binding site on the kB kinase $\beta$ for binding with an inhibitor ligand. The ligand atoms are colored by averaging the $p_{BS}(a,\mathcal{C}^B)$ (see Equation \ref{eq:bindingprob2}) values of the atoms of the kinase, and colored using the same color transfer map. Hence, ligand configuration which has a high proportion of blue, and low amount of red/orange is a better configuration (the right one in this case).}
\label{fig:bindingsite3}
\end{figure} 

A visual representation of Equation \ref{eq:bindingscore} is shown in Figure \ref{fig:bindingsite3}. For this particular example, we had a correct configuration of a ligand, that inhibits the kB kinase $\beta$, from the protein data bank model 3RZF, and a wrong configuration of the same ligand from the model 3QAD. We started with the wrong configuration and generated 1000 samples (see Figure \ref{fig:sampling}) and docked each of them using Autodock \cite{AutoDock} to the kinase. The figures show two ligand conformations. The one on the left had a poor RMSD, 3.24, with respect to the known correct configuration (and orientation) from 3RZF. The one on the right has a favorable RMSD, 1.069. Interestingly, both of them received the same score, -8.2, from AutoDock. However, when we computed the $bindingscore$ for them, the left one scored 55213, and the right one 55667, clearly showing the superiority of the latter.

\section{Results}
In this section, we detail the results of applying our QMC based UQ framework to generate Chernoff-like bounds for a set of 57 protein complexes. Additionally, we provide a protocol to determine the number of samples is required to guarantee the accuracy of the empirical certificates for specific proteins. The results clearly establish the necessity of rigorous quantification of uncertainties, and also shows that such an endeavor need not be prohibitively time consuming.

\subsection{Benchmark and experiment setup}
We applied the QMC approach of empirically bounding uncertainties of computationally evaluated QOIs to 61 crystal structures including 2 bound chains each. We took the `Rigid-body' cases of antibody-antigen, antibody-bound and enzyme complexes from the zlab benchmark 4 \cite{benchmark4}.

For each of the complexes, we applied the sampling to the receptor and the ligand (the two chains in the structure) separately, and evaluated the uncertainty measures for the calculation of surface area, volume, and components of free energy including Lennard-Jones, Coulombic, dispersion, GB and PB. We also computed the uncertainties in the binding interface area, and change of free energy. In the following subsections we explore different aspects of this analysis.

\subsection{Uncertainty of unperturbed models}
\begin{SCfigure}
 \centering
 \includegraphics[width=0.5\linewidth]{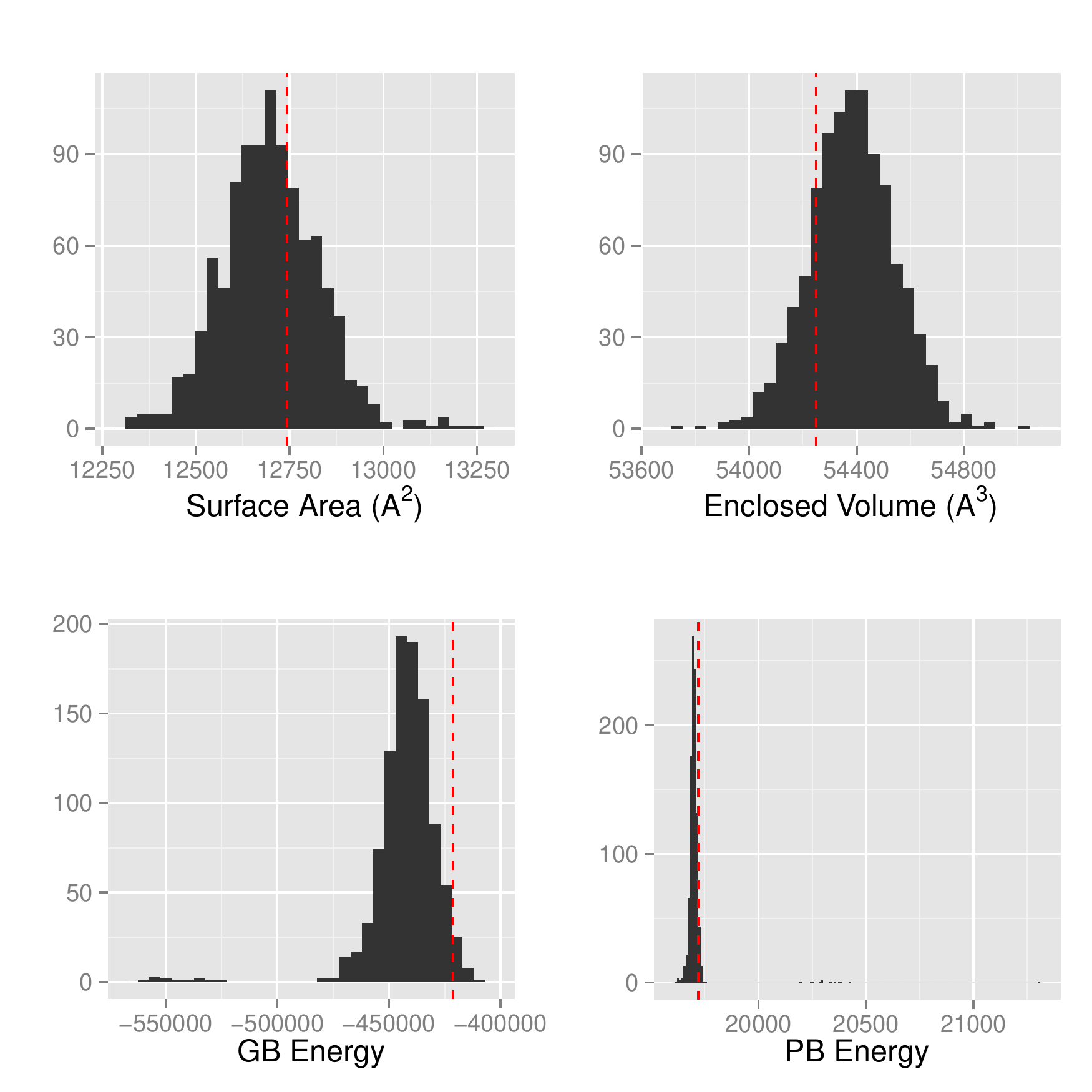}
\caption{Histogram of sampled QOIs (surface area, volume, GB energy and PB energy) for 1OPH:A. Clockwise from top left: surface area, volume, PB energy, and GB energy. The red vertical line is the value of QOI computed using the original coordinates reported in the PDB.}\label{fig:hist}
\end{SCfigure}

Figure \ref{fig:hist} shows the distribution of values computed for the sampled models for PDB structure 1OPH-chainA. The red lines in the figure shows that the original coordinates does not always provide the best estimate of the expected value of a QOI. The z-scores for these structures, with respect to the expected values and standard deviations derived from the empirical distribution, are 0.33, -0.82, 1.37, and 0.25 respectively for area, volume, GB, and PB. This emphasizes the importance of applying some form of empirical sampling to find the best representative model (one which minimizes the z-score, for instance).

\subsection{Certificates for computational models}
In the next section we discuss the likelihood of producing a large error in the calculation of QOI, due to the presence of uncertainty in the input, in terms of Chernoff-like bounds. For each model in the dataset, we computed the distribution, and then computed the probability, $\epsilon$, of a randomly sampled model having more than 0.1\%, 0.5\%, 1\%, 2\%, 3\%, 5\% and 10\% errors ($t$), where error is defined as $|x'-E[x]|/E[x]$ such that $x'$ is the value computed for a random model and $E[x]$ is the expected value (from the distribution).

Table \ref{tab:chernoff} lists Chernoff bounds as described above for the two chains of 1OPH. The rows named $\delta area(A)$ represent the quantity $|area(A+B) - area(A) - area(B)|$ computed while keeping $A$ fixed and sampling the distribution of $B$; rows named $\delta area(B)$ report the same quantity while keeping $B$ fixed and sampling the distribution of $A$.

\begin{table}[ht]
\centering
\caption{Chernoff-like bounds for the 1OPH protein. For each value of $t$, the corresponding values of $\epsilon$ are calculated from the 1000 random samples (see Equation~\ref{eqn:chern-x} for definition).}
\scriptsize

\begin{tabular}{@{}rrrrrrrr@{}}
  \hline
 $t$ & 0.001 & 0.005 & 0.01 & 0.02 & 0.03 & 0.05 & 0.1 \\ 
  \hline
area(A) & 0.911 & 0.613 & 0.326 & 0.050 & 0.012 & 0.000 & 0.000 \\ 
  area(B) & 0.911 & 0.582 & 0.306 & 0.052 & 0.003 & 0.000 & 0.000 \\ 
  $\Delta$area(A) & 0.964 & 0.823 & 0.650 & 0.358 & 0.167 & 0.031 & 0.000 \\ 
  $\Delta$area(B) & 0.973 & 0.889 & 0.771 & 0.560 & 0.375 & 0.155 & 0.005 \\ 
   \hline
vol(A) & 0.727 & 0.088 & 0.003 & 0.000 & 0.000 & 0.000 & 0.000 \\ 
  vol(B) & 0.765 & 0.174 & 0.005 & 0.000 & 0.000 & 0.000 & 0.000 \\ 
  $\Delta$vol(A) & 0.877 & 0.485 & 0.169 & 0.009 & 0.001 & 0.000 & 0.000 \\ 
  $\Delta$vol(B) & 0.895 & 0.495 & 0.176 & 0.006 & 0.001 & 0.000 & 0.000 \\ 
   \hline
LJ(A) & 0.230 & 0.012 & 0.012 & 0.012 & 0.006 & 0.001 & 0.000 \\ 
  LJ(B) & 0.281 & 0.005 & 0.004 & 0.004 & 0.003 & 0.000 & 0.000 \\ 
  $\Delta$LJ(A) & 0.636 & 0.079 & 0.034 & 0.020 & 0.015 & 0.012 & 0.001 \\ 
  $\Delta$LJ(B) & 0.967 & 0.625 & 0.210 & 0.030 & 0.020 & 0.015 & 0.001 \\ 
   \hline
CP(A) & 0.536 & 0.018 & 0.012 & 0.008 & 0.000 & 0.000 & 0.000 \\ 
  icp(B) & 0.594 & 0.013 & 0.004 & 0.000 & 0.000 & 0.000 & 0.000 \\ 
  $\Delta$CP(A) & 0.851 & 0.363 & 0.081 & 0.014 & 0.012 & 0.003 & 0.000 \\ 
  $\Delta$CP(B) & 0.828 & 0.351 & 0.070 & 0.012 & 0.012 & 0.000 & 0.000 \\ 
   \hline
GB(A) & 0.970 & 0.835 & 0.660 & 0.396 & 0.208 & 0.061 & 0.012 \\ 
  GB(B) & 0.964 & 0.817 & 0.618 & 0.310 & 0.137 & 0.019 & 0.003 \\ 
  $\Delta$GB(A) & 0.986 & 0.929 & 0.863 & 0.725 & 0.605 & 0.364 & 0.099 \\ 
  $\Delta$GB(B) & 0.982 & 0.934 & 0.866 & 0.726 & 0.601 & 0.372 & 0.094 \\ 
   \hline
PB(A) & 0.970 & 0.864 & 0.727 & 0.508 & 0.315 & 0.114 & 0.014 \\ 
  PB(B) & 0.966 & 0.838 & 0.672 & 0.403 & 0.232 & 0.047 & 0.004 \\ 
  $\Delta$PB(A) & 0.990 & 0.942 & 0.863 & 0.719 & 0.585 & 0.378 & 0.106 \\ 
  $\Delta$PB(B) & 0.984 & 0.940 & 0.870 & 0.748 & 0.640 & 0.463 & 0.164 \\ 
   \hline
\end{tabular}
\label{tab:chernoff}
\end{table}

The take-away from this table is that for most of the quantities of interest we focused on, the probability of incurring more than 5\% error is negligible (if a randomly perturbed model is used). We also note that the probability of error is higher for $\delta$ QOIs simply because the errors of individual quantities are getting propagated and amplified. Uncertainties are also higher in more complex functionals.

Please see supplement for a table in the same format which shows the average uncertainties across the entire dataset (instead of a single protein).

\subsection{Number of samples sufficient to provide statistically accurate certificates}
The results reported in the previous two subsections highlights the importance of uncertainty quantification and also shows that the mean coordinates do not always correlate well with the QOI computed from the original molecules. Once a molecule has been sampled many times, it is easy to determine the Chernoff-like bounds and quantify the uncertainty. However, it is far too costly in terms of time to generate 1000 samples every time the value of a QOI should be reported. For the next experiment, we sought to determine the minimal number of samples before the gain achieved through more samples was negligible.

For each QOI, we calculated the mean quantity for a certain number of random samples, $r$, then calculated the Chernoff-like bounds on this reduced dataset. We did the same for $s$ random samples, and computed the error between the two set of bounds (we used both $s=1000$ for correlation with the full dataset and $s=r+10$ for an incremental correlation). If the error between the these bounds was less than a given threshold, $\tau$, then we determined we had reached saturation. For our experiments, we used values of $r$ from 2 to 1000 (full dataset), values of $s$ at either 1000 or $r+10$, and a value of $\tau$ at 0.05.

An analytic solution for the average distance between two points in $d$-space has been derived \cite{hypercubeDist}, and the precomputed values for several dimensions are available online \cite{sloane2014online}. Our Chernoff-like bounds had 6 values of $t$ (dimension 6), which corresponds to an average expected distance of 0.9689. We then considered the error between two points to be the percentage distance (L2-norm) from this expected value.

\begin{SCfigure}
 \centering
\includegraphics[width=0.45\linewidth]{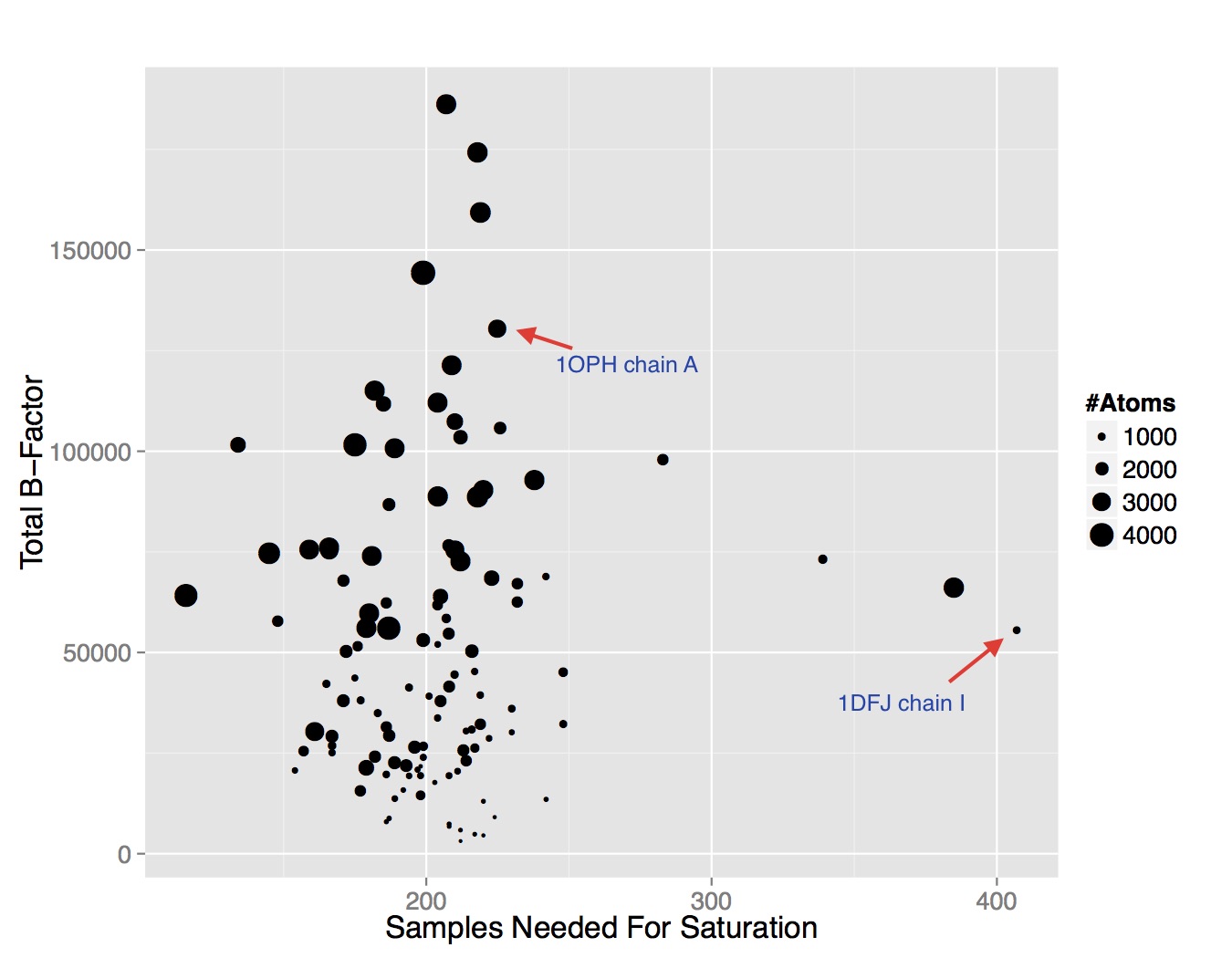}
\caption{Plot of total B-factor (a measure of both size and uncertainty) against number of samples needed before the relative error is negligible ($\tau=0.05$). For similar graphs on additional QOIs, please see supplemental data.}\label{fig:bfactor_chern}
\end{SCfigure}

Table~\ref{tab:nsamp} shows the number of samples required to reach saturation for each QOI. The variation for PB energy from the full dataset (second number) varies from 174 to 209 (2nd number in each cell), meaning that in order to calculate accurate accurate Chernoff-like bounds for the PB energy for 1OPH chain B to at most 5\% error, it is necessary to combine the result from 174 random samples. As can be seen in Figure~\ref{fig:bfactor_chern}, the number of required samples varies significantly from protein to protein. For instance, 1OPH has quite high B-factors, but converges much quicker than 1DFJ, which requires 407 samples before saturation is reached.

In practice we will not have the full 1000 samples to compare with. Instead, we must use the incremental method of sampling ($r+10$), which corresponds to the first set of numbers in Table~\ref{tab:nsamp}. These numbers are still significantly less than 1000, but are also always greater than the number of samples required for saturation against the complete dataset. Thus, we are also assured that we are within the global error bounds. For a set of figures that shows the decay in error by number of samples, please see the supplemental data.

\begin{table}[ht]
\centering
\caption{Number of samples necessary to obtain a percentage distance from the expected value less than $\tau=0.05$. Each entry consists of two numbers, the first is when compared with $x+t$ samples (where $t$ is 10), and the second is the correlation of the Chernoff-like values when compared to the entire dataset of 1000 samples. Data computed for PDB 1OPH, chains A and B. For a similar table computed on 1DFJ, see supplemental data.}\label{tab:nsamp}
\begin{tabular}{rrrrr}
  \hline
 & B & A& $\Delta$B & $\Delta$A \\ 
  \hline
area & 230/134 & 233/153 & 215/146 & 351/197 \\ 
  vol & 119/72 & 102/64 & 103/55 & 332/205 \\ 
  LJ & 79/49 & 240/168 & 315/133 & 324/192 \\ 
  CP & 79/43 & 114/62 & 93/71 & 355/213 \\ 
  GB & 281/143 & 319/202 & 300/207 & 326/211 \\ 
  PB & 287/174 & 365/209 & 357/206 & 348/196 \\ 
\hline
\end{tabular}
\end{table}

\section{Acknowledgements}
This research was supported in part by grants from NSF (OCI-1216701737 ) and NIH (R01-GM117594) and a contract (BD-4485) from Sandia National Labs.

\section{Conclusions}
In this article we have shown that even the subtle uncertainties present in high resolution x-ray structures, can lead to significant error in computational modeling. Such errors are propagated and compounded when output from one stage of modeling is fed into the next. We considered the uncertainties commonly expressed as B-factors and evaluated how it creates uncertainty in computed quantities of interest (e.g.\ surface area, van der Waals energy, solvation energy etc.). While some existing computational protocols attempts to bound the uncertainties/error due to algorithmic or numerical approximations, they do not account for the uncertainties in the input. However, our empirical study on 57 x-ray structures of bound protein complexes, showed that there significant probability ($>10$\%) of having more than 5\% error in solvation energy (GB) calculation purely due to the input uncertainties. Hence, one must rigorously account for and bound such uncertainties.

We have proposed that input uncertainties can be modeled as random variables and the uncertainty of the computed outcome (a dependent random variable) can be bounded using Chernoff-like bounds introduced by Azuma and McDiarmid. We have also shown that such bounds are also applicable when the input random variables are dependent, and show how one can theoretically bound the probability of error for Coulombic potential calculation (and any summation of distance dependent decaying kernels in general). In the future, we aim to derive similar bounds for other biophysically relevant functions.

We have also introduced an empirical quasi-Monte Carlo approximation method based on sampling the joint distribution of the input random variables to produce an ensemble of models. The ensemble is used to approximate a distribution of values for the quantity of interest. This distribution in turn can be used to bound the uncertainty of the calculation in terms of statistical certificates. A very interesting and promising outcome from application of this framework to a large set of protein structures for a wide variety of calculations showed that one typically needs fewer than 500 samples before the QMC procedure converges, hence it is quite practical to perform and report such certificates in modeling excercises. We are currently working on a graphical model of the input uncertainties, which should possibly lead to even better convergence.

Finally, we recognize that many of the uncertainties are too numerous to simply list out, and have developed several informative visualization techniques to explore and compare uncertainties in the input, intermediate states and in the final quantity of interest.

\appendix
\section{Theoretical Framework of Statistical Uncertainty Quantification}

\subsection{Martingales and McDiarmid inequality}
To prove theoretical uncertainty bounds, one often uses Chernoff-Hoeffding style bounds. However, this is useful only if the the underlying random variables are independent and we are analysing the sum of the random variables. In practical situations the random variables have dependencies. In such cases, we can still prove large deviation bounds using the theory of martingales.

\begin{define}
Let $(Z_i)_{i=1}^n$ and $(X_i)_{i=1}^n$ be a sequence of random variables on a space $\Omega$. Suppose $\E[X_i|Z_1,\ldots ,Z_{i-1}]=X_{i-1}$. Then $(X_i)$ forms a martingale with respect to $(Z_i)$.
\end{define}

We now introduce the Doob martingale. Martingales can be constructed from any random variable in the following universal way.
\begin{claim}Let $A$ and $(Z_i)$ be random variables on space $\Omega$. Then, $X_i=\E[A|Z_1,\ldots , Z_{i-1}]$ is a martingale with respect to $Z_i$. This is called the Doob martingale of $A$ with respect to $(Z_i)$.
\end{claim}

The final ingredient we need for the McDiarmid inequality is the Azuma inequality.
\begin{claim}[Azuma inequality]Let $(X_i)$ be  martingale with respect to $(Z_i)$. Suppose $|X_i-X_{i-1}| \leq c_i$. Then 
$$\Pr[|X_n-X_0|>t] \leq 2\exp(-t^2/2\sum_i c_i^2).$$
\end{claim}

We are now ready to state the McDiarmid inequality. 
\begin{claim}Let $(X_i)$ be independent random variables. Let $f:\prod_i A_i \to \mathbb{R}$ for sets $A_i$. Also, suppose that $|f(x_1,\ldots x_k,\ldots ,x_n)-f(x_1,\ldots ,x_k',\ldots , x_n)| \leq c_k$. Then, for $t>0$, 
$$\Pr[|f(\mathbf{X}-\E[f]|>t]<2\exp(-2t^2/\sum_k c_k^2).$$
\end{claim}

The weak form of the McDiarmid inequality follows directly from the Azuma inequality but we are not going to require that distinction here.

\subsection{McDiarmid inequality for summation of decaying kernels}
We are interested in bounding the uncertainty of summations over decaying kernels of the form shown below, when the variables are uncertain.

\begin{equation}
 F(A,B) = \sum_{\mathbf{x_1} \in A} \sum_{\mathbf{x_2} \in B} \sum_{k=1}^n \frac{a_k}{\|\mathbf{x_1}-\mathbf{x_2}\|^{b_k}}
\label{eq:decaykernel:complete}
\end{equation}

\noindent
where $b_k$ are non-negative constants, $a_k$ are constants, and $A$ and $B$ are two sets of points.

A single decaying kernel in the above summation is represented as

\begin{equation}
 f_{\mathbf{x_1}}(\mathbf{x_2}) = \sum_{k=1}^n \frac{a_k}{\|\mathbf{x_1} - \mathbf{x_2}\|^{b_k}} 
\label{eq:decaykernel:dd2varNterm1}
\end{equation}

\noindent
where the kernel is centered at ${\mathbf{x_1}}$ and evaluated at ${\mathbf{x_2}}$. The following result is immediate:

\begin{lem}
 For a given set of $a_k$ and $b_k$, $f_{\mathbf{x_1}}(\mathbf{x_2})$ = $f_{\mathbf{0}}(\Delta\mathbf{x})$ where $\Delta\mathbf{x} = (\mathbf{x_2}-\mathbf{x_1})$.
\end{lem}

When both $\mathbf{x_1}$ and $\mathbf{x_2}$ are uncertain such that every component $x_{1i}$ of $\mathbf{x_1}$ is uniformly sampled from the interval $[l_{1i},u_{1i}]$, and every component ${x_2}_i$ of $\mathbf{x_2}$ is uniformly sampled from the interval $[{l_2}_i,{u_2}_i]$- we can assume that every component $\Delta x_i$ of $\Delta\mathbf{ x}$ is uniformly sampled from the interval $[l_i,u_i]$ computed based on $[{l_2}_i,{u_2}_i]$ and $[{l_1}_i,{u_1}_i]$. The error of $f_{\mathbf{x_1}}(\mathbf{x_2})$ due to the uncertainty of $\mathbf{x_1}$ and $\mathbf{x_2}$ can hence be equivalently computed as the error of $f_{\mathbf{0}}(\Delta\mathbf{ x})$ due to the uncertainty of $\Delta\mathbf{ x}$. In our discussion, we shall often drop the $\Delta$ when the context does not require the distinction.

\subsubsection{Single kernel at a single point}
\label{sec:decay:1kernel}

We begin with the simplest case when the kernel is embedded in 2D (the 1D case is trivial):

\begin{equation}
 {f_1}(x,y) = \frac{a}{(x^2+y^2)^{b/2}}
\label{eq:decaykernel:2d2var1term}
\end{equation}

Assuming that $x$ and $y$ are uniformly sampled from the intervals $[l_x,u_x]$ and $[l_y,u_y]$ respectively where $l_x, l_y,u_x$ and $u_y$ are non-negative, we can define the maximum deviation due to the change of $x$ as 

${D_1}_x = \max_y |{f_1}(l_x,y) - {f_1}(u_x,y)|$

\noindent
Note that $g_1(y) = {f_1}(l_x,y) - {f_1}(u_x,y)$ is positive for $l_x < u_x$, and $\frac{d}{dy}g(y) < 0$. Hence, $g_1(y)$ is maximized when $y = l_y$. So, 

\begin{eqnarray}
{D_1}_x & = & \max_y \left|{f_1}(l_x,l_y) - {f_1}(u_x,l_y)\right| \nonumber \\
        & = & |a|\left(\frac{1}{\left(l_x^2+l_y^2\right)^{b/2}} - \frac{1}{\left(u_x^2+l_y^2\right)^{b/2}}\right)
\label{eq:decaykernel:dx2d1term}
\end{eqnarray}

\noindent
${D_1}_y$ can also be computed the same way. Using McDiarmid's theory of bounded differences, we have the following result- 

\begin{lem}
 For the decaying kernel ${f_1}$ in Equation \ref{eq:decaykernel:2d2var1term}, $\Pr[|{f_1}-E[{f_1}]| >t] \leq 2e^{\frac{-2t^2}{{D_1}_x^2+{D_1}_y^2}}$ where ${D_1}_x$ and ${D_1}_y$ are defined in Equation \ref{eq:decaykernel:dx2d1term}.
\label{lemma:decaykernel:2d2var1term}
\end{lem}

The above results can be readily extended to $d$ dimensions for the function ${f_2}$ defined below. 

\begin{equation}
 {f_2}(\mathbf{x}) = \frac{a_k}{\|\mathbf{x}\|^{b_k}} 
\label{eq:decaykernel:dd2var1term}
\end{equation}

Let, ${f_2}_i(\mathbf{x}, y)$ represent ${f_2}(\mathbf{x})$ such that the value of the $i^{th}$ component is fixed to $y$. 
So we define the maximum deviation of ${f_2}$ due to the change of one variable $x_i$ between the range $[l_i, u_i]$ as:

\begin{equation}
{D_2}_i = \max_{\mathbf{x}} {g_2}_i(\mathbf{x}) = \max_{\mathbf{x}} \left|{f_2}_i(\mathbf{x}, l_i) - {f_2}_i(\mathbf{x}, u_i)\right|
\label{eq:decaykernel:dxdd1term1}
\end{equation}

Again ${g_2}_i(\mathbf{x})$ is positive and $\frac{d}{dx_j}{g_2}(\mathbf{x}) < 0$ for all components $x_j$ of $\mathbf{x}$. Hence, ${g_2}(\mathbf{x})$ is maximized when $x_j = l_j$ for all $j$ where $l_j$ is the lowest possible value for $x_j$.

\begin{equation}
{D_2}_i = |a|\left(\frac{1}{\left(\sum_k l_k^2\right)^{b/2}} - \frac{1}{\left(u_i^2 + \sum_{k\neq i} l_k^2\right)^{b/2}}\right)
\label{eq:decaykernel:dxdd1term}
\end{equation}

\begin{lem}
 For the decaying kernel ${f_2}$ defined in Equation \ref{eq:decaykernel:dd2var1term}, $\Pr[|{f_2}-E[{f_2}]| >t] \leq 2e^{\frac{-2t^2}{\sum_i {D_2}_i^2}}$ such that ${D_2}_i$ is defined as in Equation \ref{eq:decaykernel:dxdd1term}.
\label{lemma:decaykernel:dd2var1term}
\end{lem}

Note that Lemmas \ref{lemma:decaykernel:2d2var1term} and \ref{lemma:decaykernel:dd2var1term} hold even when $a<0$ (i.e. negative).

\subsubsection{Multiple kernels at a single point}
\label{sec:decay:multikernel}

Now we extend the scope to consider functions which are expressed as a sum of $n$ decaying kernels centered at the origin.

\begin{equation}
 {f_3}(\mathbf{x}) = \sum_{k=1}^n \frac{a_k}{\|\mathbf{x}\|^{b_k}} 
\label{eq:decaykernel:dd2varNterm}
\end{equation}

Let ${f_3}^k(\mathbf{x}) = \frac{a_k}{\|\mathbf{x}\|^{b_k}}$ denote the $k_th$ decaying term in Equation \ref{eq:decaykernel:dd2varNterm}. Now, the maximum deviation will be defined similar to Equation \ref{eq:decaykernel:dxdd1term1}.

\begin{eqnarray}
{D_3}_i(\mathbf{x}) & = & \max_{\mathbf{x}} g3_i(\mathbf{x}) \nonumber \\
    & = & \max_{\mathbf{x}} \left|{f_3}_i\left(\mathbf{x}, l_i) - {f_3}_i(\mathbf{x}, u_i\right)\right| \nonumber\\
    & = & \max_{\mathbf{x}} \left|\sum_k \left({f_3}_i^k(\mathbf{x}, l_i) - {f_3}_i^k(\mathbf{x}, u_i)\right)\right| \nonumber\\
    & \leq & \max_{\mathbf{x}} \sum_k \left|\left({f_3}_i^k(\mathbf{x}, l_i) - {f_3}_i^k(\mathbf{x}, u_i)\right)\right| \nonumber\\
    & \leq & n ~ \max_k \max_{\mathbf{x}} \left|\left({f_3}_i^k(\mathbf{x}, l_i) - {f_3}_i^k(\mathbf{x}, u_i)\right)\right| \nonumber\\
    & = & n ~ \max_k {D_2}_i^k
\label{eq:decaykernel:dxddNterm}
\end{eqnarray}

\noindent 
where ${D_2}_i^k$ is defined the same way as ${D_2}_i$ in Equation \ref{eq:decaykernel:dxdd1term} for the $k^th$ kernel.

\begin{lem}
 For the sum of decaying kernel ${f_3}$ given in Equation \ref{eq:decaykernel:dd2varNterm}, $\Pr[|{f_3}(\mathbf{x})-E[{f_3}(\mathbf{x})]| >t] \leq 2e^{\frac{-2t^2}{\sum_i {D_3}_i^2(\mathbf{x})}}$ such that ${D_3}_i(\mathbf{x})$ is defined as in Equation \ref{eq:decaykernel:dxddNterm}.
\label{lemma:decaykernel:dd2varNterm}
\end{lem}

\subsubsection{Multiple kernels at multiple points}
\label{sec:decay:sumkernel}

Let us define a volumetric function in $d$ dimensions as a sum over multiple kernels defined at multiple points belonging to the set $A$ as follows.

\begin{equation}
 f_4(A, \mathbf{y}) = \sum_{\mathbf{x} \in A} \sum_{k=1}^n \frac{a_k}{\|\mathbf{x} - \mathbf{y}\|^{b_k}} 
\label{eq:decaykernel:ddMvarNterm}
\end{equation} 

Now, $f_4$ can be expressed as :

\begin{eqnarray}
 f_4(A, \mathbf{y}) & = & \sum_{\mathbf{x} \in A} {f_3}_{\mathbf{x}}(\mathbf{y}) \nonumber \\
		   & = & \sum_{\mathbf{x} \in A} {f_3}_{\mathbf{0}}(\mathbf{y}-\mathbf{x}) \nonumber \\
		   & = & \sum_{\mathbf{x} \in A} {f_3}_{\mathbf{0}}(\Delta x)
\label{eq:decaykernel:ddMvarNterm2}
\end{eqnarray} 

Since, $f_4$ is a simple summation over independent points, the result in Lemma \ref{lemma:decaykernel:ddMvarNterm} follows immediately from Lemma \ref{lemma:decaykernel:dd2varNterm}.

\begin{lem}
 For the sum of decaying kernel ${f_4}(A, \mathbf{y})$ given in Equation \ref{eq:decaykernel:ddMvarNterm}, $\Pr[|f_4(A, \mathbf{y})-E[f_4(A, \mathbf{y})]| >t] \leq 2e^{\frac{-2t^2}{\sum_{\mathbf{x} \in A}\sum_i {D_3}_i^2(\Delta\mathbf{ x})}}$ such that ${D_3}_i(\Delta\mathbf{ x})$ is defined as in Equation \ref{eq:decaykernel:dxddNterm}.
\label{lemma:decaykernel:ddMvarNterm}
\end{lem}

\subsubsection{Integral over multiple kernels at multiple points}
\label{sec:decay:integral}

Finally, we bound the uncertainties in the integral function we mentioned at the beginning of this section in Equation \ref{eq:decaykernel:complete}.

\begin{lem}
 For the sum of decaying kernel $F(A,B)$ given in Equation \ref{eq:decaykernel:ddMvarNterm}, $\Pr[|F(A,B)-E[F(A,B)]| >t] \leq 2e^{\frac{-2t^2}{\sum_{\mathbf{x_1} \in A} \sum_{\mathbf{x_2} \in A} \sum_i {D_3}_i^2(\Delta\mathbf{ x})}}$ such that ${D_3}_i(\Delta\mathbf{ x})$ is defined as in Equation \ref{eq:decaykernel:dxddNterm} and $\Delta\mathbf{ x} = (\mathbf{x_2}-\mathbf{x_1})$.
\label{lemma:decaykernel:ddMvarNtermInteg}
\end{lem}

\subsection{Extensions of Doob martingale}
In applying McDiarmid's inequality, we assumed some kind of Lipschitz nature about the composition function and the fact that the random variables are independent of each other. This lead to relatively clean bounds. However, if one wishes to analyse a very genera scenario, then we can proceed as follows. First, there is a sequence of random variables $(X_i)_{i=1}^n$ taking values in a set $A$ say. They can be dependent in any way. Consider any function $f:A^n \to \mathbb{R}$. Then, by Azuma's inequality (mentioned earlier), we can have a certificate bound of the form $$\Pr[|f(\mathbf{X})-\E[f]|>t] \leq \exp(-t^2/2\sum_{i}c_i^2),$$ where the only assumption we need is 
$$|\E_{X_{i+1},\ldots , X_n}[f(\mathbf{X})|X_1,\ldots ,X_i]$$ $$-\E_{X_{i},\ldots , X_n}[f(\mathbf{X})|X_1,\ldots , X_{i-1}]| \leq c_i.$$

Thus, the change in expectation on fixing the $i$-th random variable should not be too large. One can easily see that the conditions required for Mc Diarmid's inequality immediately imply the above hypothesis and thus the certificate bound follows. However, one can also put practically minimal restriction on the random variables and do the above computation on the amount of perturbation in the expectation, at the cost of notational aesthetics. 

We give more details now. As before there is a sequence of random variables $(X_i)_{i=1}^n$ taking values in a set $A$, say, arbitrarily dependent on each other.  Consider any function $f:A^n \to \mathbb{R}$. Define a sequence of random variables for $i=1$ to $n-1$, $$B_i=\E_{X_{i+1},\ldots , X_n}f(\mathbf{X}).$$ 
By definition, $B_i$ is a function of $X_1, \ldots , X_i$. Note that the sequence $(B_i)_{i=1}^n$ forms a martingale with respect to $X_i$.

Suppose we had $|B_{i+1}-B_i| \leq c_i$ for $i=1$ to $n-1$. Note that 
$$B_0=\E_{\mathbf{X}}f(X),$$ and $$B_n=f(\mathbf{X}).$$ 
Then, by using Azuma's inequality on the martingale sequence $(B_i)$, we get 
$$\Pr[|f(\mathbf{X})-\E[f]|>t] \leq \exp(-t^2/2\sum_{i}c_i^2).$$

Note that the only requirement we needed for the certificate bound was $|B_{i+1}-B_i| \leq c_i$. We placed no restriction on the underlying random variables. We will now reduce this to a slightly stricter, albeit easier to analyse requirement.

Suppose for every $i$, and every $x_1,\ldots , x_n, x_i'$, we have 
$$\left|\E_{X_i|X_{i+1,\ldots , X_n}}f(\mathbf{X})-f(x_1,\ldots , x_i',x_{i+1},\ldots , x_n)\right| \leq c_i.$$

Then, 
\begin{eqnarray*}
&&B_{i-1}-B_i\\
&=&\E_{X_i,\ldots , X_n}f(\mathbf{X})-\E_{X_{i+1},\ldots , X_n}f(\mathbf{X})\\
&=&\E_{X_{i+1},\ldots , X_n}\E_{X_i|X_{i+1},\ldots , X_n}f(\mathbf{X})\\
&&-f(x_1,\ldots , x_i',x_{i+1},\ldots , x_n)
\end{eqnarray*}
which is bounded by $c_i$ by assumption. Therefore, we can then use Azuma's inequality on the above martingale.

We highlight this with a simple illustration. Consider the single kernel model at a single point. This will illustrate the main point. We will again consider the $2$ dimensional kernel defined by 
$$f(x,y)=\frac{a}{(x^2+y^2)^{b/2}}.$$

Let $(X,Y)$ be a variable following a joint distribution. Note that we do not require independence between $X$ and $Y$. Define
$$B_0(x,y)=f(x,y),$$ $$B_1(x)=\E_Y[f(x,Y)],$$ $$B_2=\E_{X,Y}[f(X,Y)].$$

Now, for most reasonable distributions, it is relatively straightforward to prove $$|\E_{Y|X=x}[f(x',Y)]-f(x,Y)| \leq c.$$

This immediately implies that both 
$$|B_1-B_0|, |B_2-B_1| \leq c.$$ 
Using Azuma's inequality, we conclude the required large deviation bound certificate.

\section{Details of Free Energy Model}
The Gibbs model of free energy of a molecule is defined as $E = E_{MM} + E_{sol} - TS$ where $E_{MM}$ is the molecular mechanical energy representing the atom-atom interactions among atoms of the molecule, and $E_{sol}$ is the solvation energy representing the interaction of the molecule with the solvent (usually water with some charged ions), $T$ is the temperature and $S$ is the entropy. The change of free energy upon binding or the binding free energy is defined as $\delta(A,B) = E(A+B) - (E(A) + E(B))$, i.e.\ the difference of the total free energies before and after binding. 

The molecular mechanical energy $E_{MM}$ consists of both bonded and non-bonded interaction terms. The bonded energy terms  (Equation \ref{eq:EMMbonded}) measure the energy required to deviate from an optimal bonded position (for example, changing the length of a bond).

\begin{equation}
\label{eq:EMMbonded}
\Ed = \sum _{bond~length~(d)}k_{d} ( d - d_{eq} )^{2},~~
\Etheta = \sum _{bond~angle~(\theta)}k_{\theta } ( \theta -\theta _{eq} )^{2},~~
\Ephi = \sum _{torsion~(\varphi)}k_{\varphi } ( 1-\cos [ n( \varphi - \varphi _{eq} ) ] ), 
\end{equation}

The non-bonded terms (Equation \ref{eq:EMMnonbonded}) include the van der Waals energy (also called Lennard-Jones potential) $\Evdw$ representing short range attraction-repulsion where $r_{ij}$ is the distance between two atoms, and $a_{ij}$ and $b_{ij}$ are two weights which have been determined based on quantum mechanics for different types of atom-atom pairs. The Coulombic term  $\Ecoul$ is the long range electrostatic interaction between two molecules. This term is dependent on the charges $q$ as well the distance dependent dielectric $\epsilon{\left( r_{ij} \right)}$ of the solvent. While computing the binding free energy of two rigid molecules in vacuum, these are the most relevant quantities, and hence are often used as primary scoring terms in docking and homology modeling.

\begin{equation}
\label{eq:EMMnonbonded}
\Evdw = \sum_{i}{ \sum_{j > i}{\left( { a_{ij} \over { r_{ij}^{12} } } - { b_{ij} \over { r_{ij}^6 } } \right) } }, ~~ 
\Ecoul = \sum_{i}{ \sum_{j > i}{ {q_{i} q_{j}} \over { \epsilon{\left( r_{ij} \right)} r_{ij} } } }
\end{equation}

Solvation energy is often expressed as $\Gsol = \Gcav + \Gvdw + \Gpol$ where $\Gcav$ depends on the volume of the protein and the exposed surface area and $\Gvdw$ is the Van der Waals interaction between exposed atoms and solvent atoms, and a polariation energy term $E_{\textrm{pol}}$. The polarization energy has the form $\Gpol = \frac{1}{2}\int(\phi(\bx)- \phi_{\rm gas})\rho(\bx) d \bx$ where $\rho$ is the charge density and $\phi$ and $\phi_{\rm gas}$ are the electrostatic potential for the molecule in solution and in a gas, respectively. This energy can be computed two ways: by modeling the potential with the {\em Poisson-Boltzmann (PB) equation} \cite{Sharp1990}, or by approximating the energy with the {\em Generalized Born (GB) model} \cite{Still1990}. 

We use a simplified solution to a boundary integral formulation of the PB equation \cite{Juffer:BEM}
{\footnotesize
\begin{equation}\label{eq:pbeq}
\phi(\bx) - \phi_{\rm gas}(\bx) = \int_\Gamma \left(\frac{\epsilon_E}{\epsilon_I}\frac{\p G_\kappa(\bx,\by)}{\p \bvn(\by)} - \frac{\p G_0(\bx,\by)}{\p\bvn(\by)}\right) \phi(\by) d \by \nonumber 
+  \int_\Gamma \left( G_0(\bx,\by) - G_\kappa(\bx,\by) \right)\frac{\p \phi(\by)}{\p \bvn (\by)} d\by
\vspace{-.07in}
\end{equation}
}
where $G_\kappa(\bx,\by)$ is the fundamental solution of the PB equation and $\phi_{\rm gas}$ is the potential of the molecule in a uniform dielectric.

For, GB we use the approximation given in \cite{Still90} as-
\vspace{-0.2cm}
\begin{equation}
\label{eq:Epol}
E_{\textrm{pol}} = - {\tau \over 2} \sum_{i, j}{{q_{i} q_{j}} / {\sqrt{r_{ij}^{2} + R_{i} R_{j} e^{-{ {r_{ij}^{2}} \over {4 R_{i} R_{j}} }}}}},
\end{equation}
\noindent
where $\tau = 1 - {1 \over {\epsilon}}$, and $R_{i}$ is the effective Born radius of atom $i$.


\section{Details of Free Energy Model}

\appendix

\section*{Supplementary Data and Figures}
\begin{table}[ht]
\centering
\caption{Chernoff bounds for all proteins, averaged.}
\begin{tabular}{rrrrrrrr}
  \hline
 $t$ & 0.001 & 0.005 & 0.01 & 0.02 & 0.03 & 0.05 & 0.1 \\
  \hline
area & 0.9097 & 0.5748 & 0.2821 & 0.0593 & 0.0177 & 0.0041 & 0.0001 \\ 
  $\Delta$area & 0.9104 & 0.5781 & 0.2875 & 0.0659 & 0.0241 & 0.0094 & 0.0029 \\ 
  volume & 0.7671 & 0.1908 & 0.0377 & 0.0065 & 0.0026 & 0.0007 & 0.0000 \\ 
  $\Delta$volume & 0.7681 & 0.1925 & 0.0380 & 0.0064 & 0.0025 & 0.0007 & 0.0000 \\ 
  ilj & 0.3571 & 0.0967 & 0.0897 & 0.0787 & 0.0676 & 0.0491 & 0.0259 \\ 
  $\Delta$ilj & 0.3619 & 0.0979 & 0.0895 & 0.0785 & 0.0672 & 0.0489 & 0.0257 \\ 
  icp & 0.6290 & 0.1073 & 0.0585 & 0.0325 & 0.0224 & 0.0091 & 0.0036 \\ 
  $\Delta$icp & 0.6301 & 0.1075 & 0.0581 & 0.0322 & 0.0222 & 0.0090 & 0.0036 \\ 
  gb & 0.9625 & 0.8171 & 0.6487 & 0.3906 & 0.2346 & 0.1000 & 0.0286 \\ 
  $\Delta$gb & 0.9627 & 0.8181 & 0.6506 & 0.3939 & 0.2384 & 0.1032 & 0.0296 \\ 
  pb & 0.9690 & 0.8461 & 0.7080 & 0.4716 & 0.3031 & 0.1250 & 0.0226 \\ 
  $\Delta$pb & 0.9693 & 0.8473 & 0.7103 & 0.4759 & 0.3086 & 0.1318 & 0.0296 \\ 
   \hline
\end{tabular}
\end{table}

\begin{table}[ht]
\caption{Table of number of samples to reach saturation for 1DFJ, chains E and I. Corresponding to Table 2 in the main paper.}
\centering
\begin{tabular}{rrrrr}
  \hline
 & E & I & $\Delta$E & $\Delta$I \\ 
  \hline
area & 265/157 & 471/295 & 421/240 & 337/199 \\ 
  vol & 156/69 & 385/273 & 428/257 & 343/218 \\ 
  LJ & 319/223 & 462/361 & 507/361 & 363/206 \\ 
  CP & 235/138 & 652/401 & 607/509 & 353/214 \\ 
  GB & 341/219 & 335/193 & 296/217 & 303/227 \\ 
  PB & 360/235 & 546/375 & 567/430 & 338/225 \\ 
   \hline
\end{tabular}
\end{table}

\begin{figure}
\centering
 \includegraphics[width=0.7\linewidth]{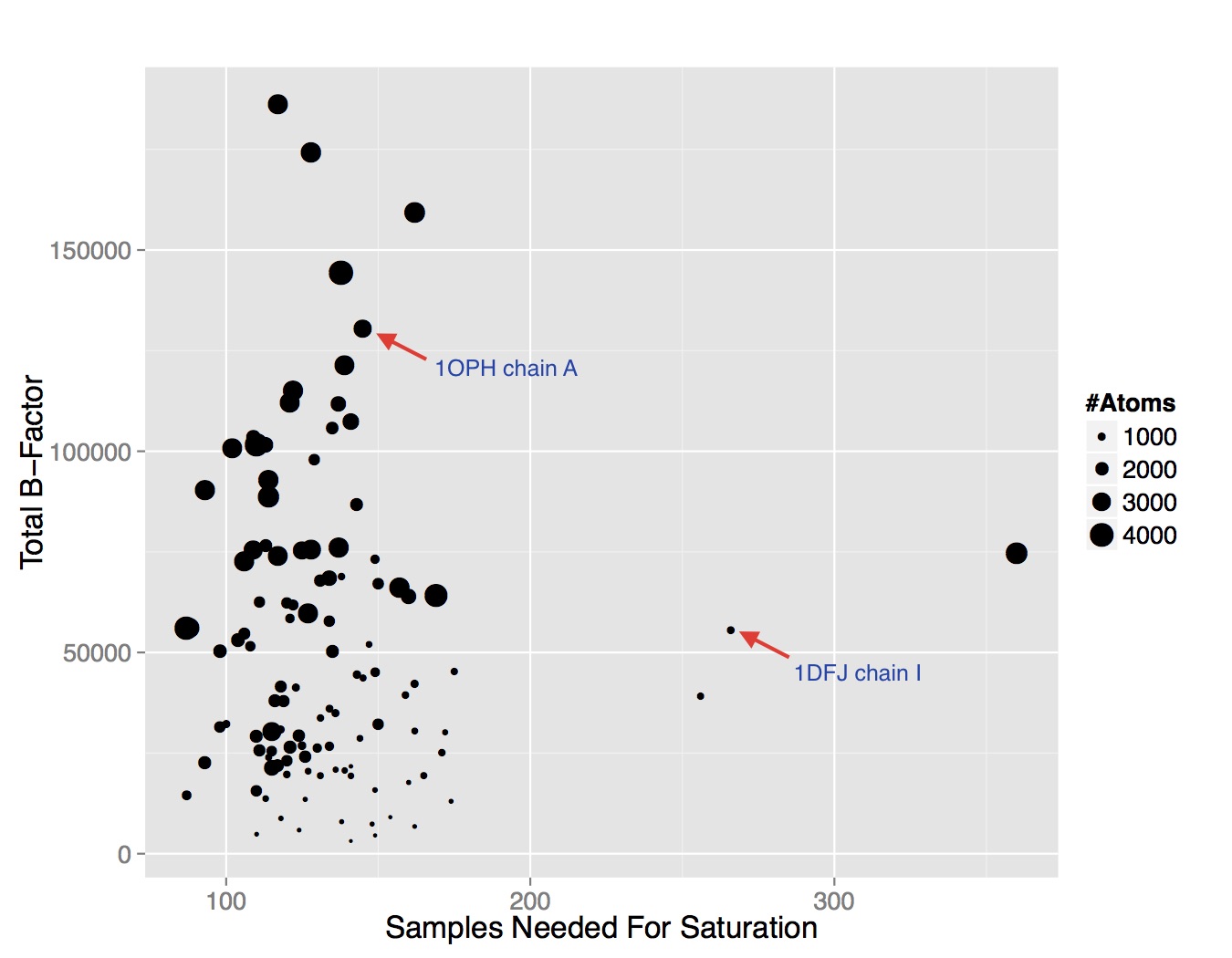}
\caption{Plot of total B-factor (a measure of both size and uncertainty) against number of samples needed before the relative error is negligible ($\tau=0.05$). QOI is surface area.}
\end{figure}
\begin{figure}
\centering
\includegraphics[width=0.7\linewidth]{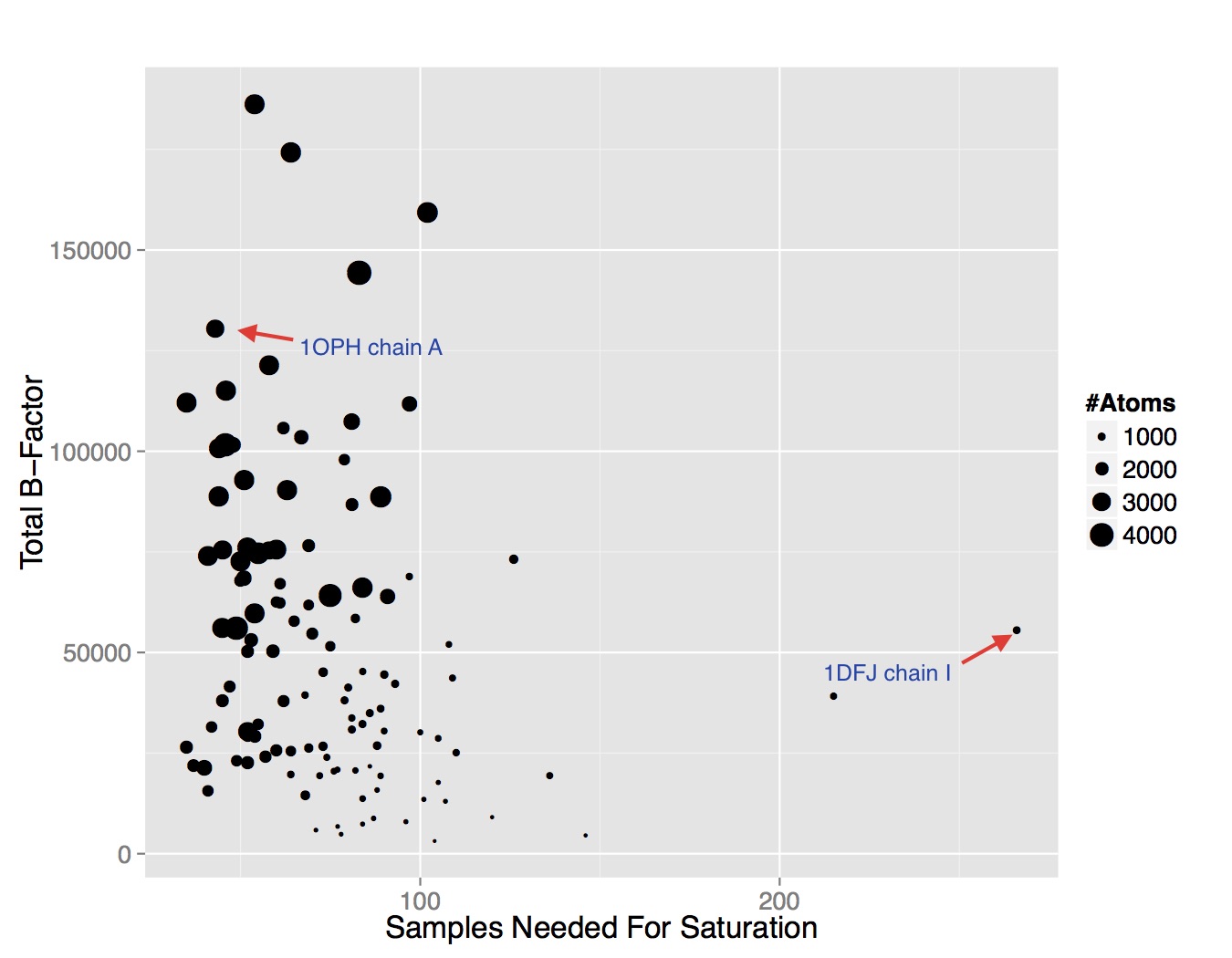}
 \caption{Plot of total B-factor (a measure of both size and uncertainty) against number of samples needed before the relative error is negligible ($\tau=0.05$). QOI is enclosed volume.}
\end{figure}

\begin{figure}
\centering
\includegraphics[width=0.7\linewidth]{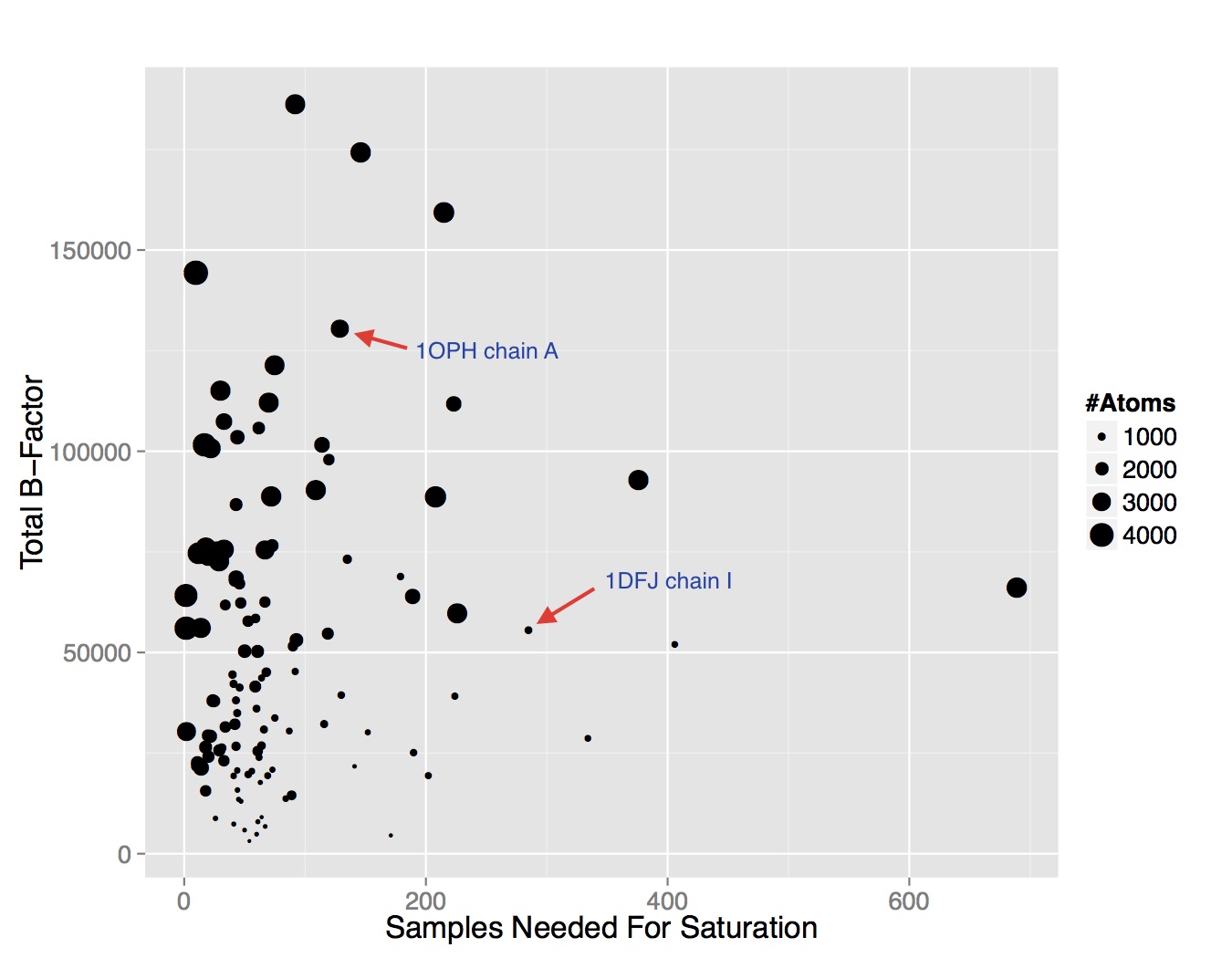}
 \caption{Plot of total B-factor (a measure of both size and uncertainty) against number of samples needed before the relative error is negligible ($\tau=0.05$). QOI is LJ potential.}
\end{figure}

\begin{figure}
\centering
\includegraphics[width=0.7\linewidth]{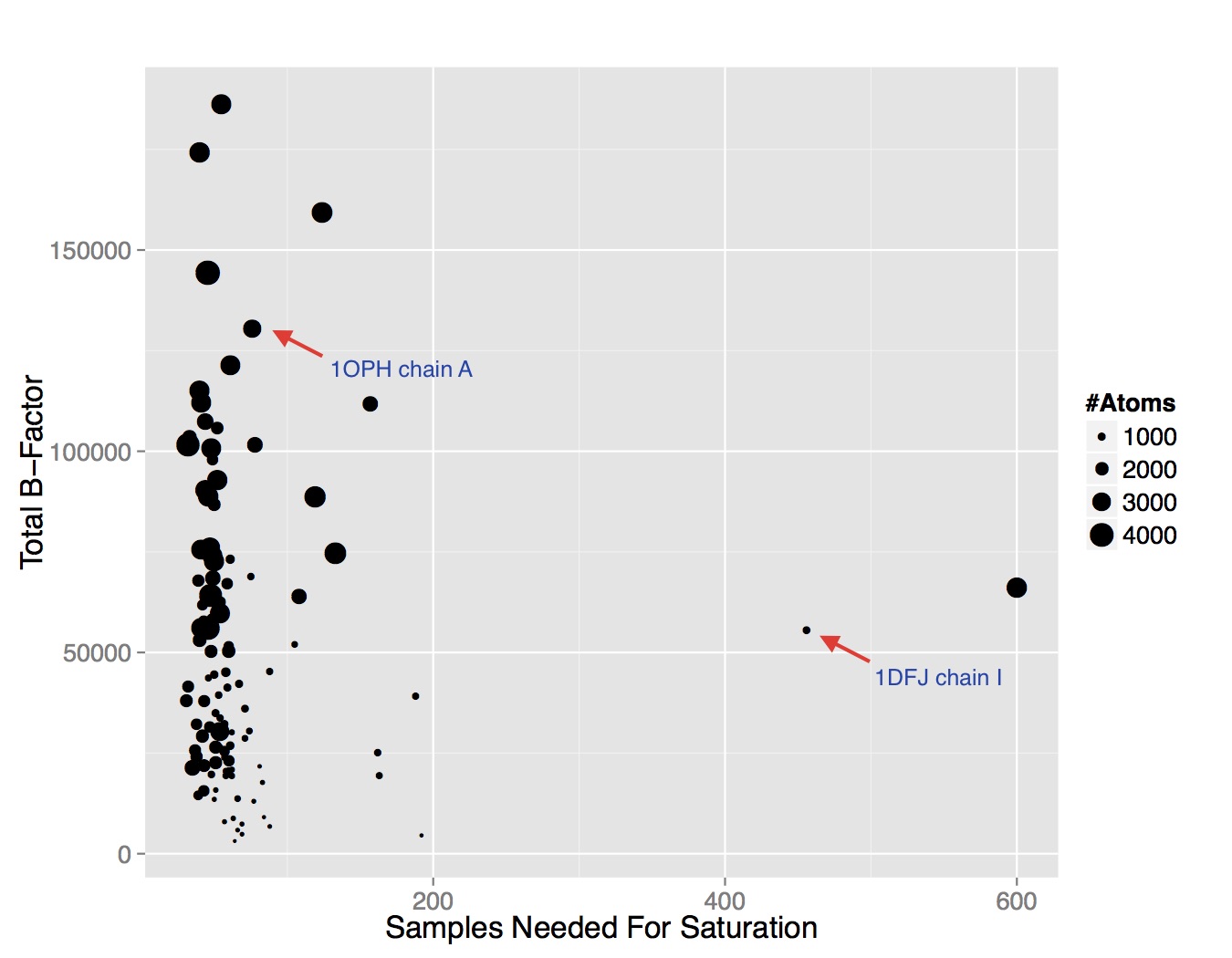}
 \caption{Plot of total B-factor (a measure of both size and uncertainty) against number of samples needed before the relative error is negligible ($\tau=0.05$). QOI is Coulombic potential.}
\end{figure}

\begin{figure}
\centering
\includegraphics[width=0.7\linewidth]{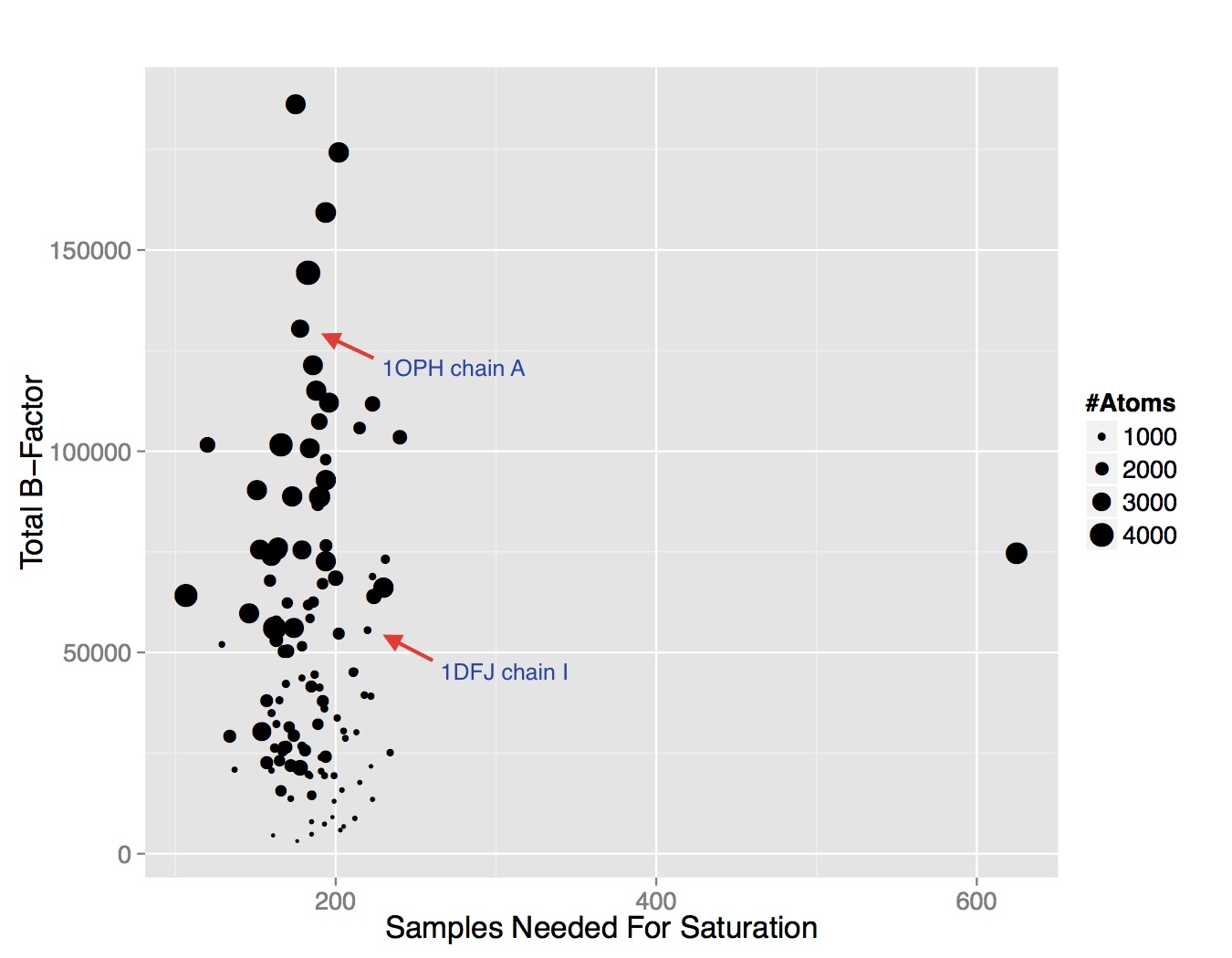}
 \caption{Plot of total B-factor (a measure of both size and uncertainty) against number of samples needed before the relative error is negligible ($\tau=0.05$). QOI is GB energy.}
\end{figure}

\begin{figure}
\centering
 \includegraphics[width=0.75\linewidth]{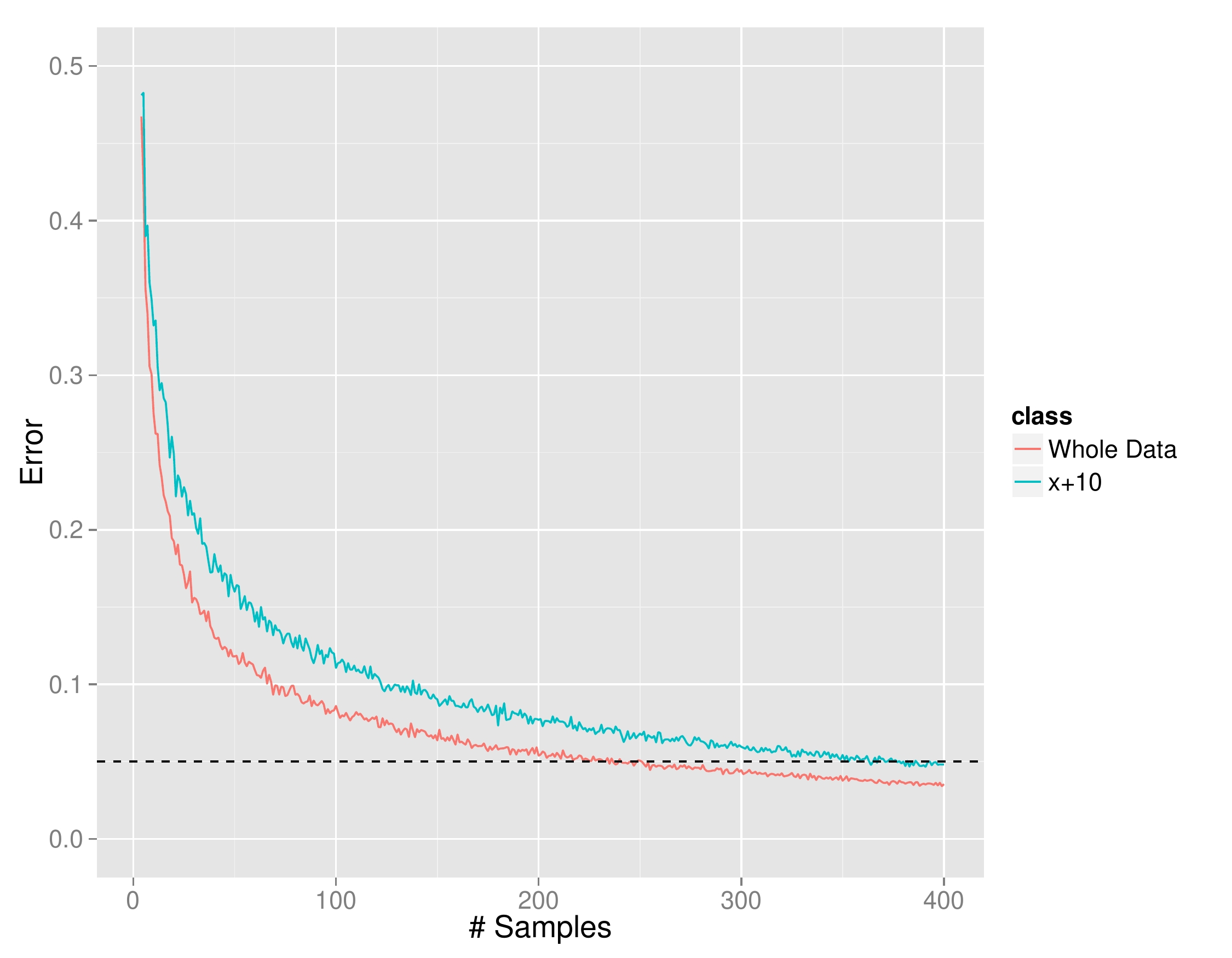}
 \includegraphics[width=0.75\linewidth]{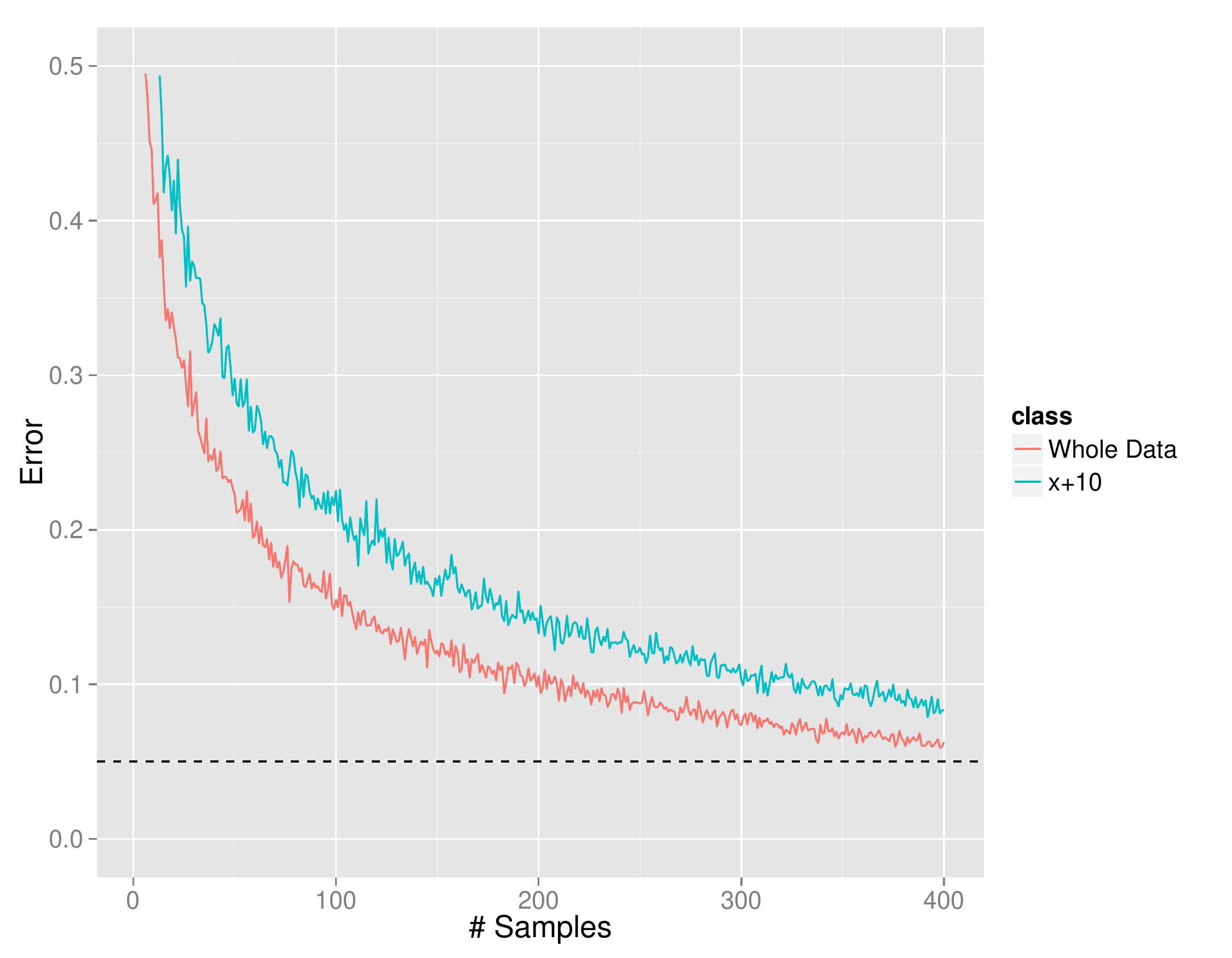}
\caption{Plots showing the rate of convergence for PB energy certificates in terms of the error metric defined in the ``Number of samples sufficient...'' section in the main text. Red line shows the error against the Chernoff-like bounds from the entire dataset; blue line shows the incremental error. Top graph is for 1OPH chain A and bottom is 1DFJ chain I.}
\end{figure}

\begin{figure}
 \centering
 \includegraphics[width=0.9\linewidth]{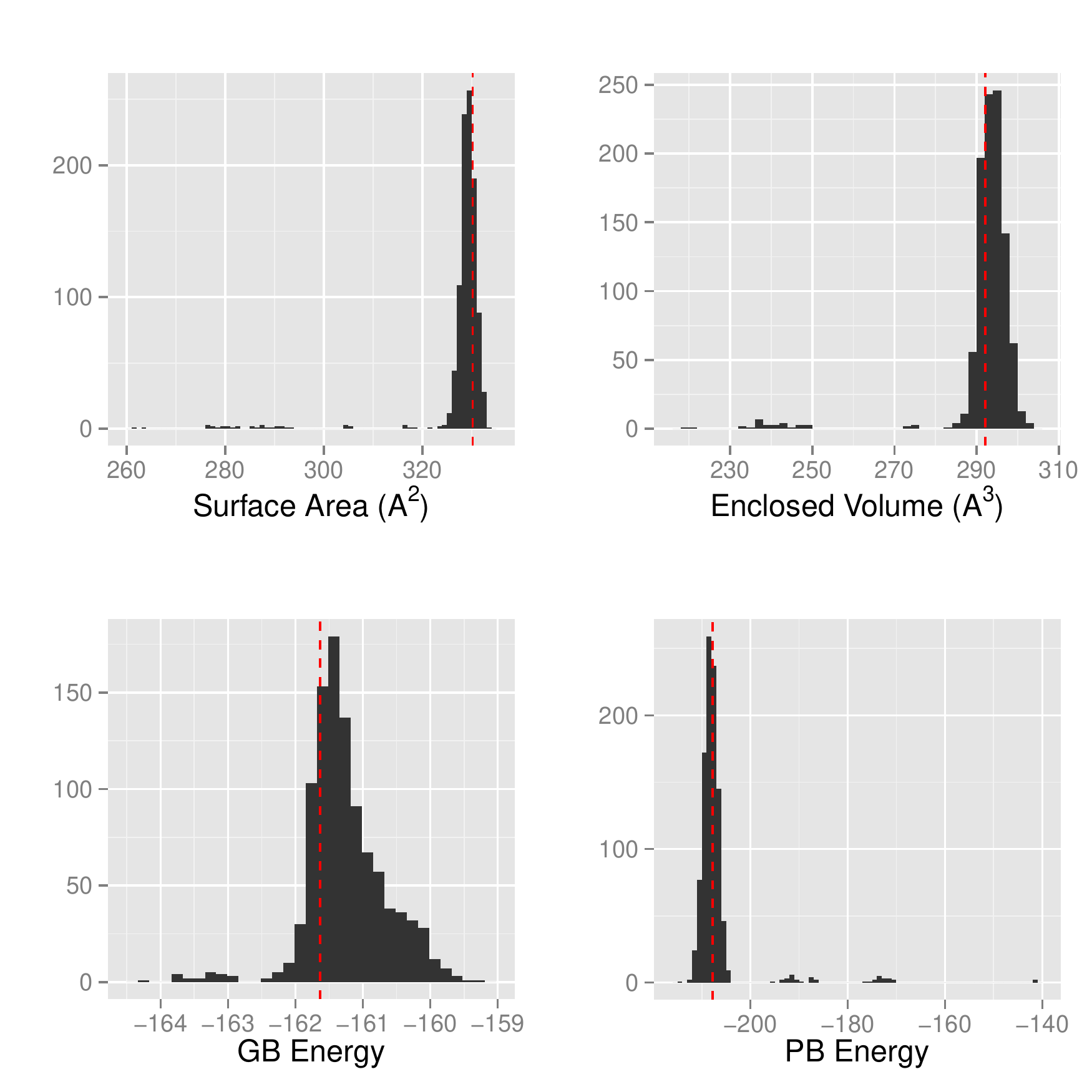}
\caption{Histogram of sampled QOIs for 3QAD ligand. Clockwise from top left: surface area, volume, PB energy, and GB energy. The red vertical line is the value of QOI computed using the original coordintates reported in the PDB.}
\end{figure}

\begin{figure}
\centering
 \includegraphics[width=0.75\linewidth]{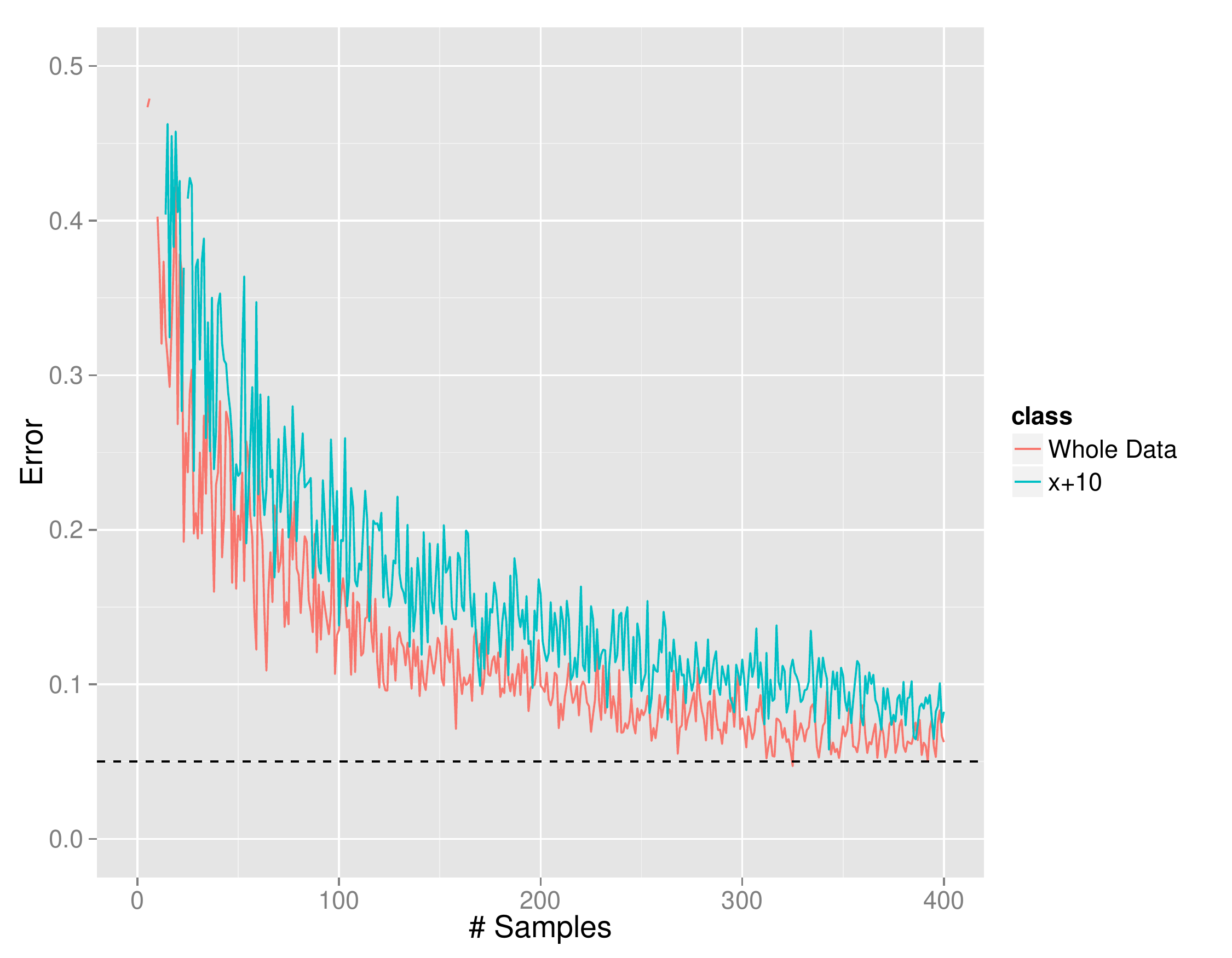}
\caption{Plots showing the rate of convergence for PB energy certificates in terms of the error metric defined in the ``Number of samples sufficient...'' section in the main text. Red line shows the error against the Chernoff-like bounds from the entire dataset; blue line shows the incremental error. Graph is for the sampled ligand from 3QAD. Individual samples behave more erratically because minor perturbations in a single torsion angle can produce a drastically different molecule.}
\end{figure}

\end{document}